\documentclass[prl,aps,twocolumn,groupedaddress,showpacs,floatfix,preprintnumbers]{revtex4}
\usepackage{amsmath,amssymb,multirow,epsfig,bm,color}
\usepackage{hyperref}
\usepackage{url}
\usepackage{color}

\newcommand{\fig}[1]{Fig.~\ref{#1}}

\newcommand{\beq}{\begin{equation}}
\newcommand{\eeq}{\end{equation}}
\newcommand{\bea}{\begin{eqnarray}}
\newcommand{\eea}{\end{eqnarray}}

\newcommand{\fm}{\;\textrm{fm}}
\newcommand{\MeV}{\;\textrm{MeV}}

\newcommand{\ssec}[1]{\emph{#1}.---}

\begin{document}

\title{
Novel Role of Superfluidity in Low-Energy Nuclear Reactions
}

\author{Piotr Magierski$^{1,2}$, Kazuyuki Sekizawa$^{1}$, Gabriel Wlaz\l{}owski$^{1,2}$}

\affiliation{$^1$Faculty of Physics, Warsaw University of Technology, Ulica Koszykowa 75, 00-662 Warsaw, Poland}
\affiliation{$^2$Department of Physics, University of Washington, Seattle, Washington 98195--1560, USA}

\email{magiersk@if.pw.edu.pl, sekizawa@if.pw.edu.pl, gabrielw@if.pw.edu.pl}
 
\begin{abstract} 
We demonstrate, within symmetry unrestricted time-dependent density
functional theory, the existence of new effects in low-energy nuclear
reactions which originate from superfluidity. The dynamics of the pairing
field induces solitonic excitations in the colliding nuclear systems,
leading to qualitative changes in the reaction dynamics. The solitonic
excitation prevents collective energy dissipation and effectively suppresses
fusion cross section. We demonstrate how the variations of the total
kinetic energy of the fragments can be traced back to the energy stored
in the superfluid junction of colliding nuclei. Both contact time and
scattering angle in non-central collisions are significantly affected.
The modification of the fusion cross section and possibilities for
its experimental detection are discussed. 
\end{abstract}


\maketitle

\begin{figure}[t]
   \begin{center}
   \includegraphics[width=0.48\textwidth, trim=0 70 0 30, clip]{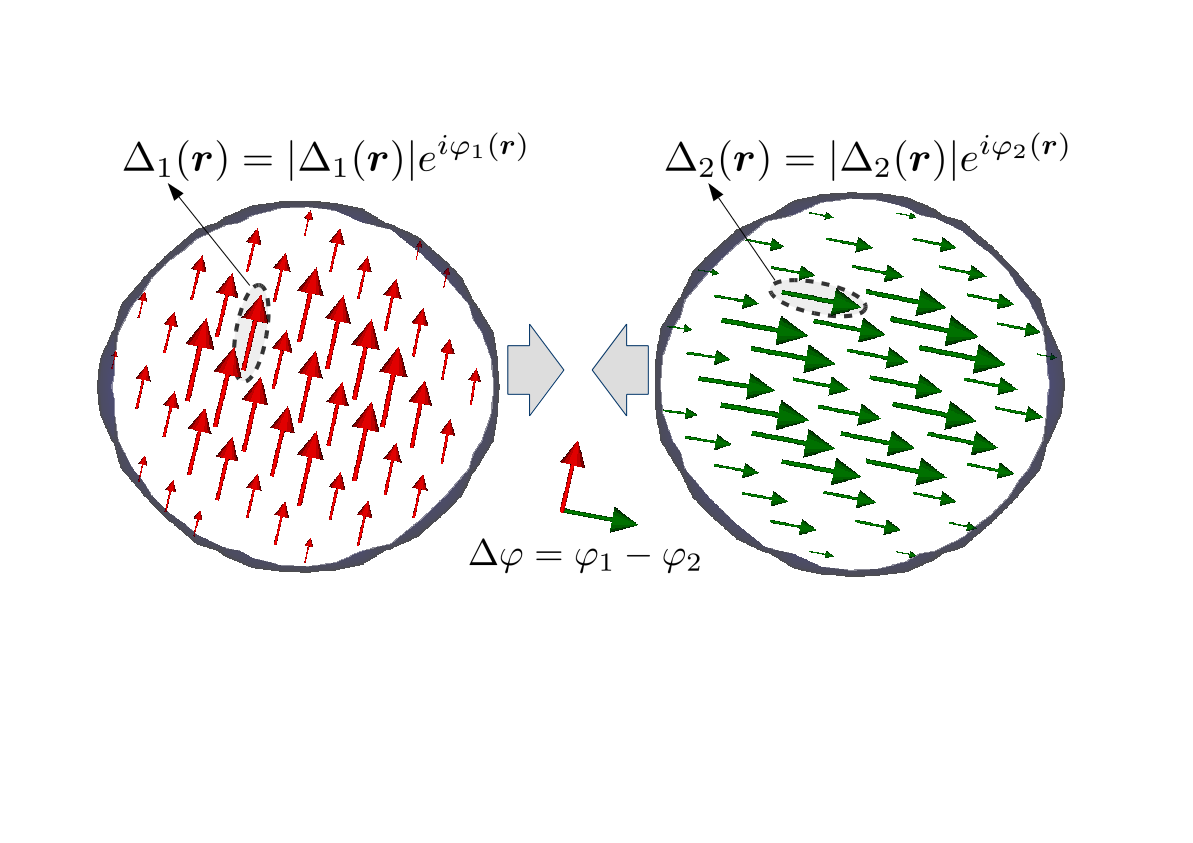}
   \end{center}\vspace{-3mm}
   \caption{(Color online)
   Schematic picture of the situation we examine in the
   present Letter: a collision of two superfluid nuclei
   with different phases of the pairing fields. Each disc represents a cross
   section of a nucleus. The arrows inside the nucleus
   indicate the paring field $\Delta_i(\bm{r})$ where length of the arrow
   indicates its absolute value $|\Delta_i(\bm{r})|$, while direction indicates its phase $\varphi_i(\bm{r})$ ($i=1,2$). In the ground state, the phase is uniform across each nucleus $\varphi_i(\bm{r})=\varphi_i$ and the phase difference $\Delta\varphi$ ($\equiv\varphi_1-\varphi_2$) is well defined. We will show how the phase difference affects the reaction dynamics.
   }
   \label{fig:idea}
\end{figure}

\ssec{Introduction} 
Dynamics of the pairing field during the nuclear reactions has
rarely been investigated to date, although it is well-known that 
the static pairing field is crucial for the description of the
atomic nuclei, both in the ground state as well as in excited states
(see, \textit{e.g.}, \cite{ring,shimizu,bender,dean,hashimoto2013} and references therein).
The reason is twofold: first, it is believed that the pairing field
dynamics will produce only small corrections to the commonly accepted
picture of low-energy nuclear reactions; second, the proper treatment
of the pairing field dynamics requires to use more advanced approaches
resulting in rapid increase of computational complexity. On the other hand,
it is well known that the pairing correlations give rise to abundant fascinating phenomena, like topological excitations, observed with great details in superfluid helium~\cite{VortexHe} or ultracold atomic gases~\cite{MIT1,MIT2}.
For example in experiments with ultracold atomic gases, where two clouds of atomic Bose-Einstein Condensates (BEC) are forced to merge, the interface between the two BECs may lose its superfluid character (solitonic excitation). This excitation is unstable and decays through quantum vortices~\cite{PRL__2014,PRA__2015}. In this paper, we investigate the possibility of creating similar excitations in nuclear reactions, see Fig.~\ref{fig:idea}.

The pairing field in nuclear systems is small in a sense that
the ratio of its magnitude to the Fermi energy does not exceed $5\%$.
It implies that BCS treatment is regarded as a justified approximation
and the size of the Cooper pair is of the same order as the size of
a heavy nucleus. Although the pairing field is small as compared to,
\textit{e.g.}, the unitary Fermi gas~\cite{becbcs}, it is important
for the proper description of the nuclear systems:
while it smears out shell effects responsible for static deformations, it also enables large-amplitude collective motion which otherwise would be strongly damped.
Therefore the description of nuclear fission requires to take into account 
superfluidity as one of crucial ingredients~\cite{bertsch1980,barranco1990,bertsch1994}.
Recently, it has been pointed out that dynamic excitations of the pairing field,
which is absent in the static treatment, affect significantly the induced fission process leading to much
longer fission timescales than predicted by other simplified approaches~\cite{PRL__2016}.

The pairing field $\Delta(\bm{r})$ can be regarded as an order
parameter that specifies whether the nucleus is superfluid or not~\cite{BrinkBroglia}.
The order parameter belongs to U(1) universality class
and it can be decomposed as $\Delta(\bm{r})=|\Delta(\bm{r})|e^{i\varphi(\bm{r})}$.
In the ground state the phase is uniform across the nucleus,
and it can be absorbed by the gauge transformation.
Then the only relevant quantity is its absolute value $|\Delta(\bm{r})|$
which is on the order of $1\MeV$. The situation is different when two
superfluid nuclei collide. Then the relative phase $\Delta\varphi$
between two pairing fields is well defined (see Fig.~\ref{fig:idea})
and cannot be removed by the gauge transformation. 
This difference will trigger various excitation modes of the pairing
field as well as  the particle flow between colliding nuclei. Although
the phases of the pairing fields are not controlled in nuclear experiments,
they will affect the reaction outcomes in an averaged way. The consequences 
of this effect turn out to be significant and are discussed in this Letter. 

\ssec{Collision of two superfluid nuclei} 
Let us first focus on the energy scale of the possible effect which
may appear during a collision of two superfluid nuclei at a fixed pairing 
phase difference $\Delta\varphi$. One would naively expect that it is governed
by the pairing energy which is proportional (for protons or neutrons) to
$\frac{1}{2} g(\varepsilon_F) |\Delta|^{2}$, where $g(\varepsilon_{F})$
represents the density of states per one spin projection at the Fermi level,
and $\Delta$ is the pairing gap. Such quantity for nuclei is on the order of
MeV, and thus one may infer that the possible effects would be too weak
to be observed in nuclear reactions. However, this is not the case since
during the collision a junction between two superfluids is created, where 
the phase varies rapidly. The energy stored in the junction
depends both on the phase difference and the size of the junction. One may
estimate the energy of the junction from phenomenological theory of superfluids, namely the Ginzburg-Landau (G-L) approach:
\begin{equation}
E_{j}=\frac{S}{L}\frac{\hbar^{2}}{2m}n_{s}\sin^{2}\frac{\Delta\varphi}{2},
\label{eq:E_junction}
\end{equation}
where $S$ is the area of the junction, $L$ is the length scale 
over which the phase varies, and $n_{s}$ is the superfluid density
(for derivation, see~\cite{SM}). Note that neither the pairing energy,
nor the pairing gap enters this formula explicitly. For a collision of 
two heavy nuclei at energies close to the Coulomb barrier,
one can show that the energy stored in the junction can vary by 
several tens of MeV depending on the phase difference~\cite{SM}.
Such a drastic energy change may
significantly alter the dynamics of the collision. Clearly in order
to determine those quantities in Eq.~(\ref{eq:E_junction}) ($S,L,n_s$)
one needs to perform microscopic simulations, since they are in general
dependent on the actual reaction dynamics.
Note that the situation described here is markedly different from the Josephson effect encountered in solids, ultra-cold atomic gases or heavy ion collisions~\cite{dietrich1, dietrich2,dietrich3, sorensen, hashimoto2016}. The Josephson effect involves tunneling between weakly-coupled pairing condensates. Here we focus on the strong-coupling limit: the nature of the junction is entirely different even though its decay will also involve a Josephson-like current. In this Letter, we show that the associated pairing field dynamics has a significant impact on the fusion cross section and the total kinetic energy (TKE) of the fragments.


\ssec{TDSLDA for nuclear reactions} 
Presently, the most accurate microscopic approaches to the dynamics
of superfluid systems are based on the density functional theory~\cite{Oliveira,Wacker}.
Here we utilize an
approximated formulation known as time-dependent superfluid local 
density approximation (TDSLDA), which is formally equivalent to the
time-dependent Hartree-Fock-Bogoliubov theory (TDHFB). The approach has been proved to be 
very accurate for describing dynamics of strongly correlated fermionic systems, like ultracold
atomic gases~\cite{PRL__2009,Science__2011,LNP__2012,PRL__2012,ARNPS__2013,PRL__2014,PRA__2015}
and nuclear systems~\cite{PRC__2011,PRL__2015,PRL__2016,Mag2016}.
We solve the TDSLDA equations numerically on a 3D spatial lattice
(without any symmetry restrictions) with periodic boundary conditions.
We use a box of size $80\;{\rm fm}\times25\;{\rm fm}\times25\;{\rm fm}$ 
for head-on collisions and $80\;{\rm fm}\times60\;{\rm fm}\times25\;{\rm fm}$ 
for non-central collisions. The lattice spacing is set to $1.25$~fm. 
For the energy density functional, we use FaNDF$^0$ functional~\cite{Fayans1,Fayans2}
without the spin-orbit term.

Although it is well known that the spin-orbit interaction is 
crucial for a proper description of nuclear static properties
as well as energy dissipation in low-energy nuclear reactions,
it does not induce qualitative change in the pairing field dynamics.
In this Letter, we investigate possible impact of the phase difference
on the reaction dynamics and address the following questions: what
observables are affected by the phase difference and for each affected
quantity what is the predicted size of the effect? In order to answer
these questions one needs to set correctly the scales of the problem,
which are determined in the present context by the average
magnitude of the pairing gap and the ratio of the coherence length to 
the size of the system. None of the meaningful scales in our problem is affected
by the spin-orbit interaction. However, in order to provide quantitative results
that can be compared directly with experimental data, one needs to
perform calculations with a full nuclear density functional. We defer
these extremely numerically expensive studies to future works.
This simplification allows us to construct a highly efficient solver
of the TDSLDA equations (for details, see supplemental material
of~\cite{VortexPinning}). Nevertheless, the problem is still numerically
demanding and requires usage of supercomputers. 
Very recently, the first attempt has been reported
in \cite{hashimoto2016}, where the effects of the phase difference in head-on collisions of ${}^{20}$O+${}^{20}$O were investigated based on TDHFB including the spin-orbit contribution. In case of reactions with light systems, the impact of the phase difference on various observables was found to be very small~\cite{hashimoto2016,arXiv:1702.00069}. 

One may rise a question regarding the adequacy of the description 
of the finite system using the theoretical framework admitting the 
broken particle-number symmetry. It gives rise to the Nambu-Goldstone
(NG) modes related to the rotation of the phase of the pairing field \cite{nambu,goldstone}. 
The phase can be traced back to the phase of the Cooper-pair wave
function, which can be defined as the eigenfunction corresponding 
to the dominant eigenvalue of the two-body density matrix, and thus,
is independent of a particular approximation in the treatment of the
pairing correlations. The particle-number projected (symmetry-restored)
wave function would imply averaging over the phase. The natural question is
whether this averaging needs to be performed before the collision. 
The answer to this question is related to the timescale of the 
associated NG mode, which is governed by the nuclear chemical potentials~\cite{Nakatsukasa2016}.
Since phases of both projectile and target nuclei rotate during the
time evolution, what matters is the difference of the periods of the
phase rotations. If it is long enough, as compared to the collision time, 
the use of the framework with broken
particle-number symmetry is validated~\cite{Bulgac}. In the case
of nuclear collision it is determined by the difference of (one nucleon) separation
energies of the projectile and target nuclei $\Delta S=|S_{1}-S_{2}|$.
Thus, the description will be valid if one limits to the collision of
nuclei whose difference between the separation energies does not exceed
$1\MeV$ that leads $T=\frac{2\pi\hbar}{\Delta S} > 1200\fm/c$ which is
longer than the collision time. The most clean case corresponds to the
symmetric collision where the phase difference does not depend on time.

\begin{figure}[t]
\includegraphics[width=1.0\columnwidth, trim=110 407 100 393, clip]{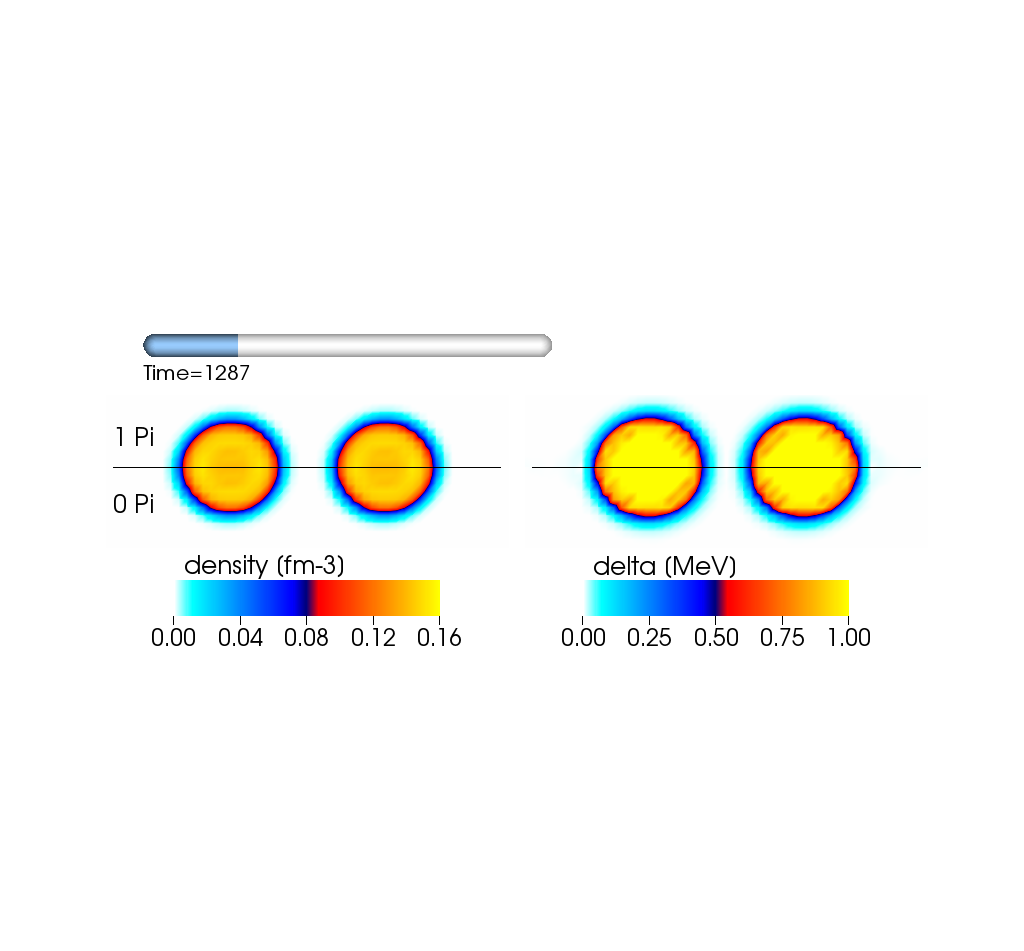}\\
\includegraphics[width=1.0\columnwidth, trim=110 407 100 393, clip]{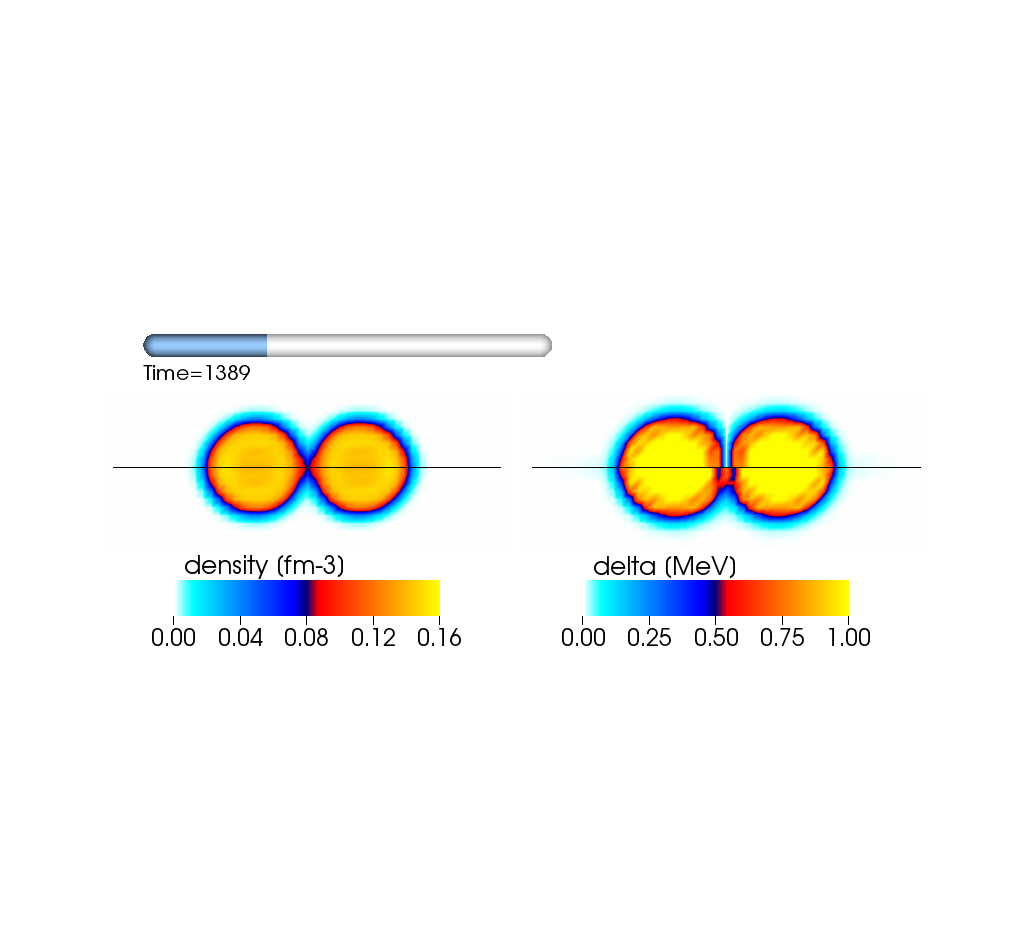}\\
\includegraphics[width=1.0\columnwidth, trim=110 407 100 393, clip]{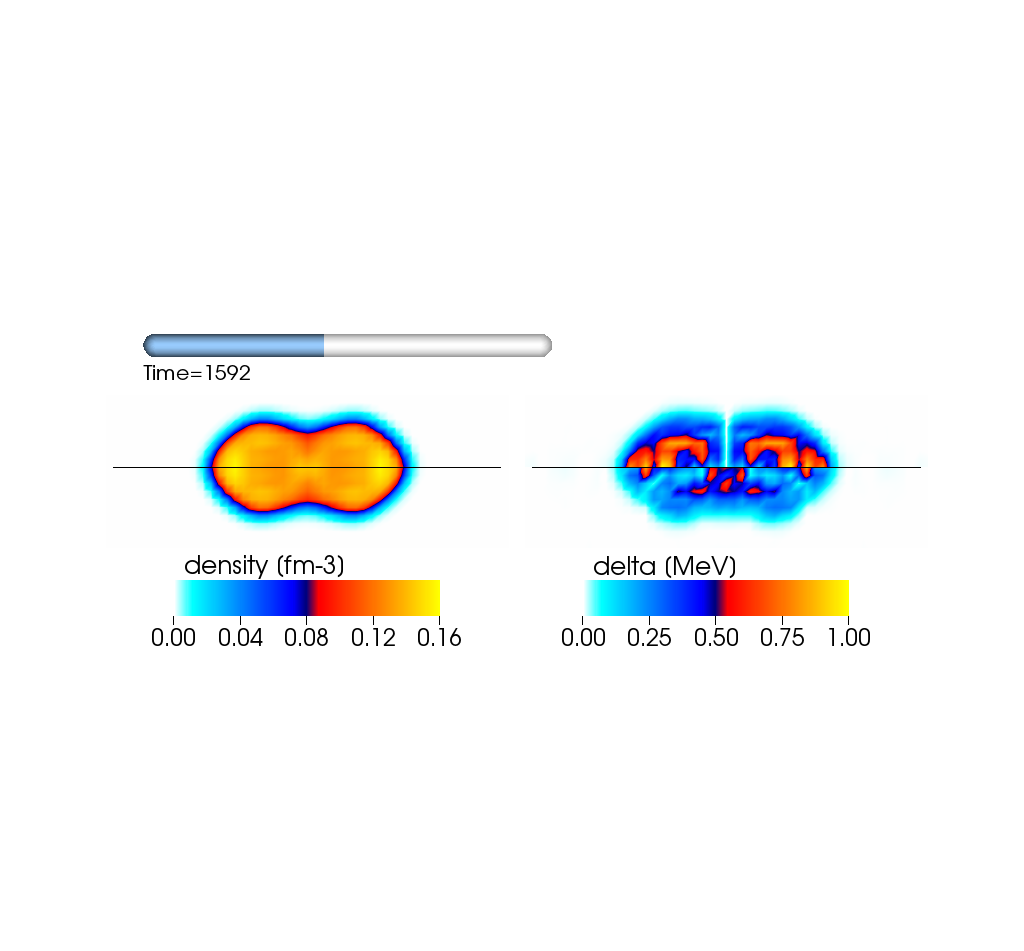}\\
\includegraphics[width=1.0\columnwidth, trim=110 407 100 393, clip]{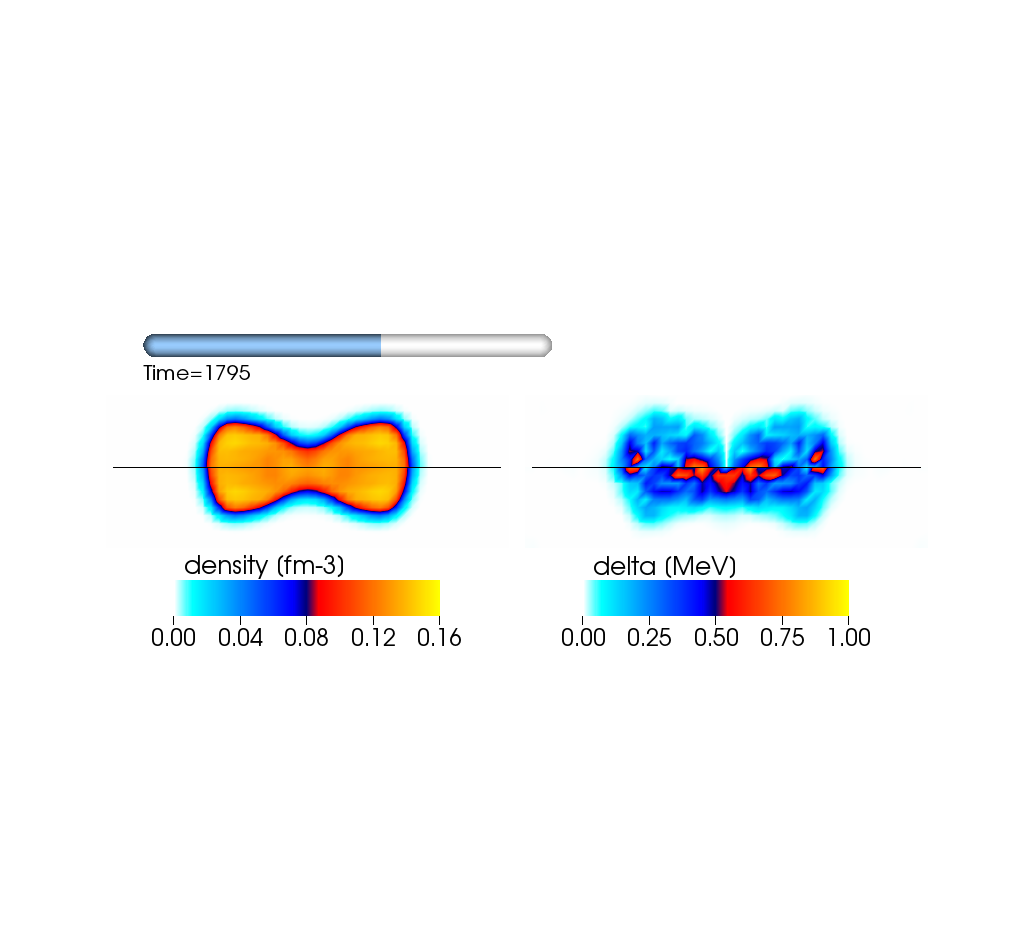}\\
\includegraphics[width=1.0\columnwidth, trim=110 305 100 393, clip]{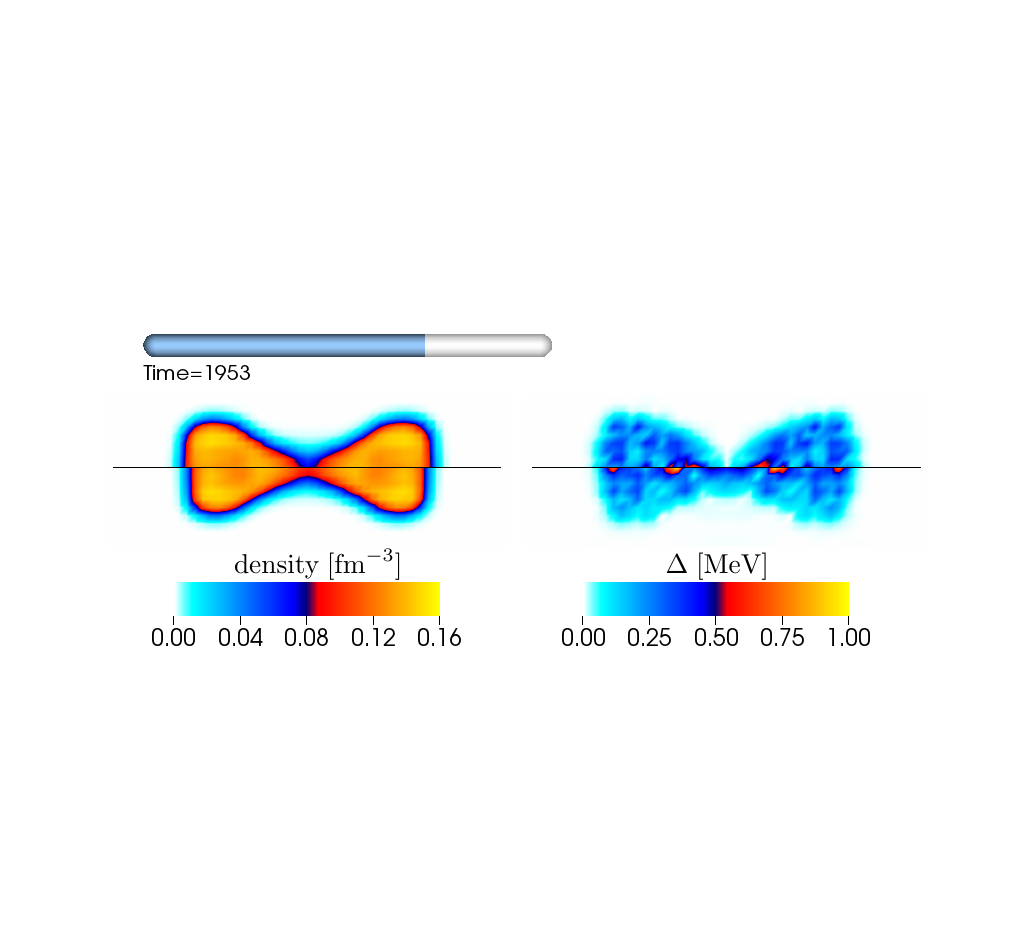}
\caption{ (Color online) Snapshots from the collision of $^{240}$Pu+$^{240}$Pu for two extreme values of the relative phase differences ($\Delta\varphi=0$ and $\pi$)
at the energy $E \simeq 1.1V_{\rm Bass}$, where $V_{\rm Bass}$ represents the phenomenological fusion barrier~\cite{Bass1974}.
Left panels show the total density distribution, whereas the right panels show the neutron paring field of two colliding nuclei. Top half of each panel corresponds to the phase difference $\Delta\varphi=\pi$ case, while bottom half corresponds to the case without phase difference $\Delta\varphi=0$. Contact time is about $550$--$600\fm/c$ depending on the phase difference. 
For movies and plots showing the phase evolution see~\cite{SM}.
\label{fig:Pu_collision}}
\end{figure}

\begin{figure}[t]
   \begin{center}
   \includegraphics[width=\columnwidth]{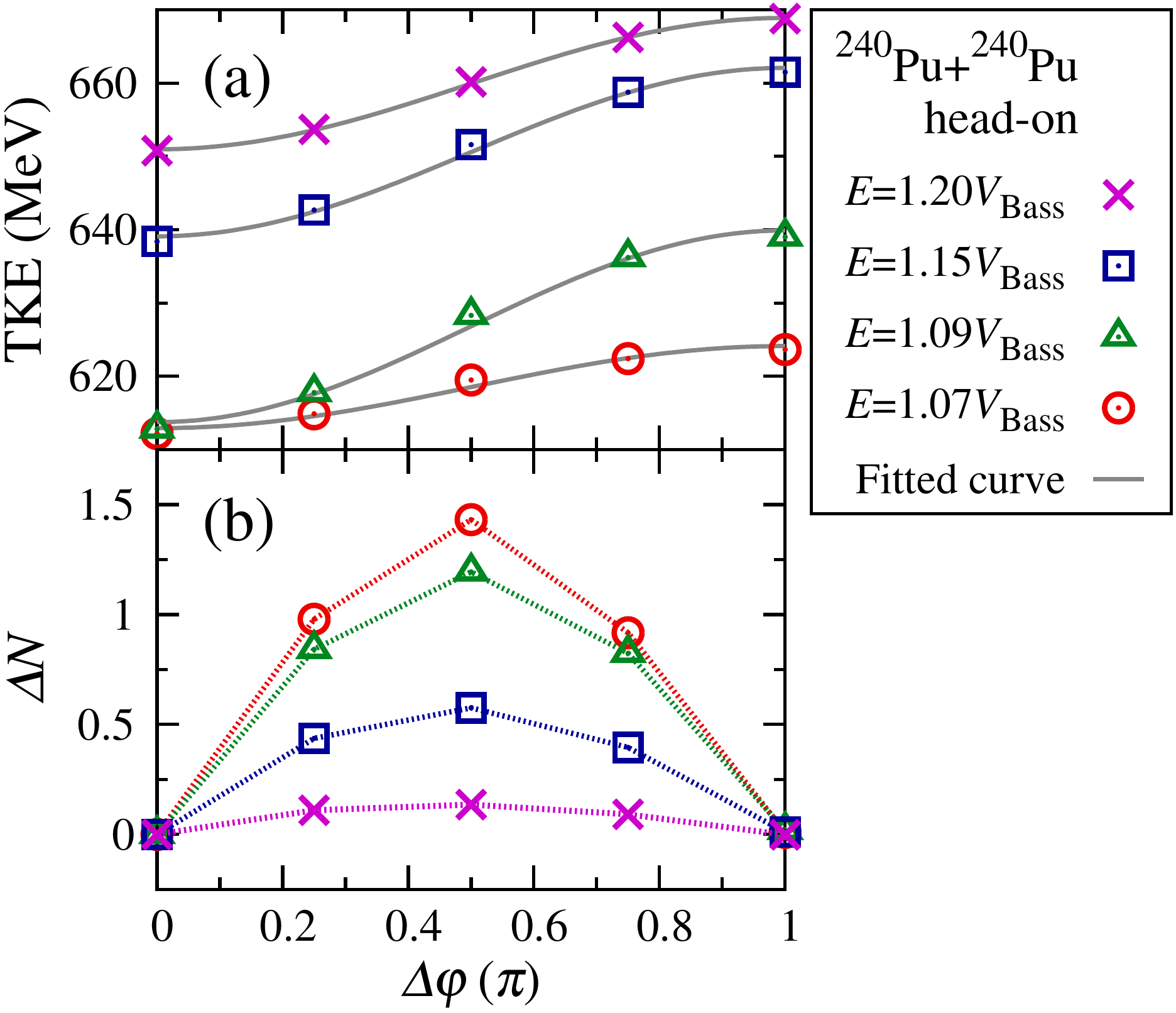}
   \end{center}\vspace{-3mm}
   \caption{(Color online)
   Results of the TDSLDA simulations for $^{240}$Pu+$^{240}$Pu 
   head-on collisions at various collision energies.
   (a): Total kinetic energy (TKE) of the outgoing fragments is shown. 
   Line shows fit to the data by a formula $\alpha+\beta\sin^{2}\frac{\Delta\varphi}{2}$ 
   with respect to parameters $\alpha$ and $\beta$.
   (b): The average number of transferred neutrons from the left nucleus to the right nucleus due to the Josephson current is shown.
   The horizontal axis is the relative pairing phase $\Delta\varphi$.
   Note that change of TKE has different phase dependence, and cannot be explained by the Josephson effect.  
   }
   \label{fig:Pu_junction}
\end{figure}

\ssec{Kinetic energy and Josephson current} 
As a first example, let us consider symmetric collisions of two heavy nuclei,
$^{240}$Pu (since the spin-orbit term is neglected the
nucleus does not exhibit a prolate deformation).
In such a case two nuclei do not fuse and reseparate shortly after collision. 
In Fig.~\ref{fig:Pu_collision},
we show pairing fields and densities of the colliding nuclei at various times
in two extreme cases, $\Delta\varphi=0$ and $\Delta\varphi=\pi$.
It is clearly
visible that in the $\Delta\varphi=\pi$ case a narrow solitonic structure 
is created, \textit{i.e.} inside the structure the order parameter vanishes, the density is suppressed and the phase changes rapidly from one value to another
when one crosses the structure. 
It stays there until the composite system splits. This produces
a significant impact on resulting TKE of the fragments. In Fig.~\ref{fig:Pu_junction}~(a),
we show the TKE as a function of the relative phase for various collision energies. 
The TKE clearly shows the $\sin^{2}\frac{\Delta\varphi}{2}$ pattern (gray
solid curves), which exactly recovers the dependence of the energy of the
junction given by the G-L approach, Eq.~(\ref{eq:E_junction}). The dominating
contribution comes from the neutron pairing field. The contribution from
the proton pairing field is less than $30\%$ of the neutron effect, due
to Coulomb repulsion~\cite{SM}. These results indicate that the phase
difference hinders the energy transfer from the relative motion to internal
degrees of freedom. We emphasize that the observed change of TKE cannot be attributed to the Josephson effect. For example, for extreme cases $\Delta\varphi=0$ and $\Delta\varphi=\pi$, there is no Josephson current (as it scales like $\sin\Delta \varphi$) while dynamics of the reaction is altered.

The situation becomes qualitatively different when the energy is further
increased. Namely, at energies about $30\%$ above the barrier, the departure
from this simple pattern is observed. It corresponds to the energies at 
which a third light fragment is generated~\cite{SM}.
The appearance of the third light fragment in the quasifission
process is understood as a consequence of the density and charge excesses
in the neck region~\cite{golabek}. However the solitonic excitation effectively
reduces the density in the neck region. Consequently, for the energy range 
$1.3V_{\rm Bass} < E < 1.5V_{\rm Bass}$ the number of fragments depends on
the phase difference and smaller phase differences favour the creation of 
the third fragment~\cite{SM}. For $E > 1.5V_{\rm Bass}$ the ternary quasifission
is observed for all phase differences.

The stability of the solitonic excitation described here depends on the possibility of phase transfer between the pairing fields of the colliding nuclei, which manifests itself as particle transfer. Even though the reaction is symmetric, it can cause nucleon transfer from one nucleus to the other. Indeed, after reseparation the fragments are not symmetric. However, the amount of nucleon transfer does not exceed $1.5$ for neutrons and $0.5$ for protons during the collision (see, Fig.~\ref{fig:Pu_junction}~(b) and \cite{SM} for more details). This result is consistent with earlier studies~\cite{dietrich1, dietrich2, dietrich3, sorensen, hashimoto2016}. Note that this particle transfer resembles a Josephson current, even though the solitonic excitation itself has an entirely different origin.

\begin{figure}[t]
   \begin{center}
   \includegraphics[width=0.35\textwidth]{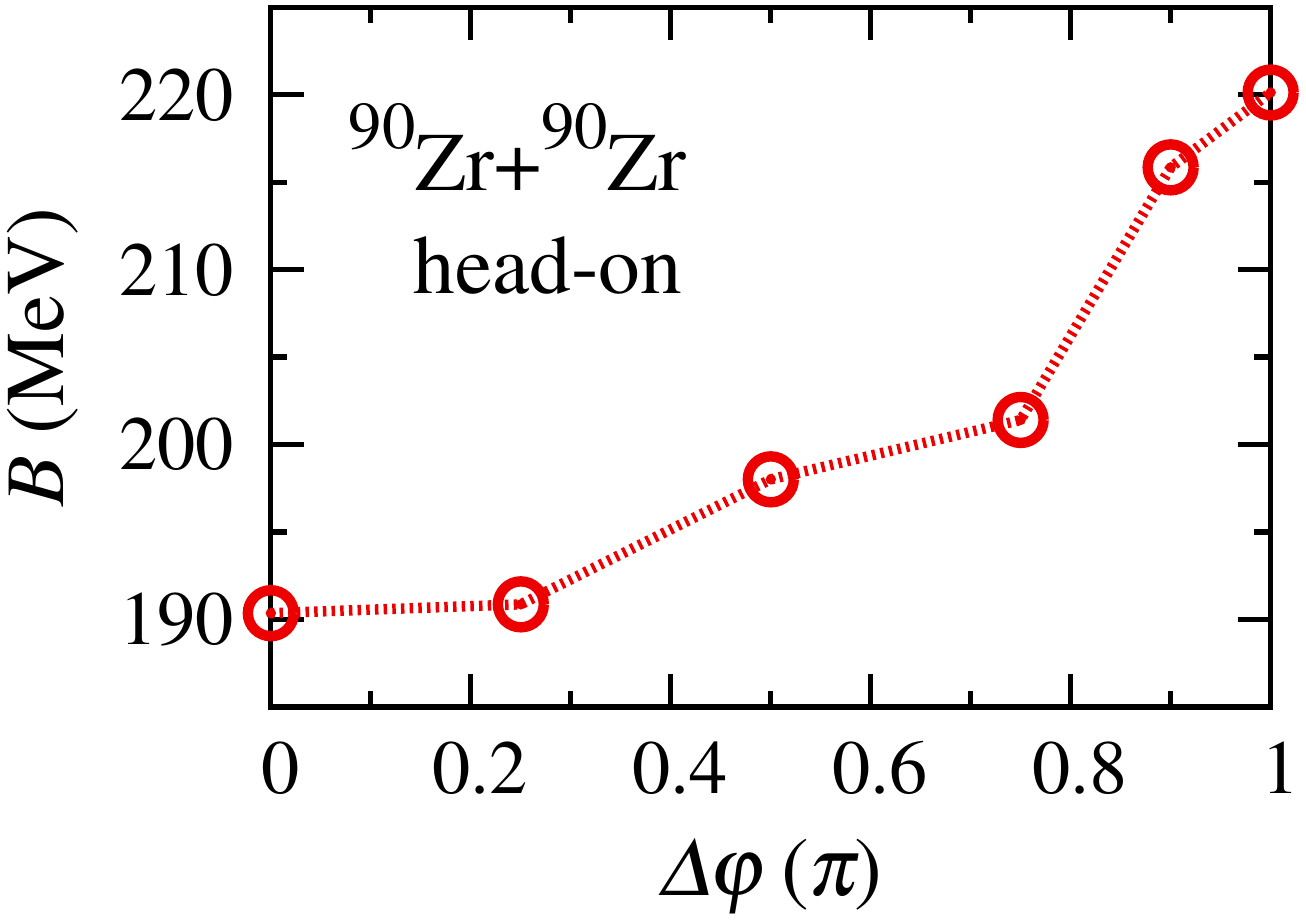}
   \end{center}\vspace{-3mm}
   \caption{(Color online)
   Results of the TDSLDA simulations for $^{90}$Zr+$^{90}$Zr head-on collisions.
   Fusion threshold energy $B$ is shown as a function of the relative pairing 
   phase $\Delta\varphi$. For this reaction the barrier height is $V_{\rm Bass}\simeq192\MeV$. 
   The phase difference prevents fusion for energies up to $15\%$ above the barrier. 
   }
   \label{fig:Zr_fusion}
\end{figure}

\ssec{Energy threshold for fusion}
Results for the heavy system indicate that the phase difference effectively
works as a potential barrier, and consequently it will affect the fusion cross
section. In order to investigate this issue, we examine collisions of two
medium mass nuclei, $^{90}$Zr, that can fuse.
Note that when the spin-orbit
term is dropped this is an open-shell nucleus for neutrons and thus neutrons
are superfluid, whereas protons occupy a closed shell. In Fig.~\ref{fig:Zr_fusion},
we show the minimum energy required for the system to merge in head-on collisions
and stay in contact for times longer than $12\,000\fm/c$ (40~zs). The results
clearly demonstrate that fusion reaction is effectively hindered by the
dynamic excitations of the pairing field. The energy threshold as a function
of the angle does not have $\sin^2\frac{\Delta\varphi}{2}$ dependence, since
we consider now collisions varying both the phase difference and the collision energy. 

The fusion hindrance phenomenon associated with pairing dynamics may likely be observed by studies of the fusion
cross section for symmetric systems at the vicinity of the barrier, in a similar 
way to experimental detection of the so-called extra-push energy~\cite{Sahm1985,liang2012},
which is the energy introduced by Swiatecki to explain the experimental fusion cross 
sections for collisions of medium and heavy nuclei at energies above the Coulomb barrier
\cite{swiatecki1982,swiatecki1982a,donangelo1986}.
As a good candidate we suggest symmetric collisions of different Zr isotopes.
For these reactions the extra-push energy is negligible. $^{90}$Zr is neutron
magic ($N=50$) and the pairing correlations are absent.
As the neutron number increases 
neutrons become superfluid  which hinders the fusion reaction. Based on our 
results the extra energy for fusion is expected to be about 
$E_{\rm{extra}}=\frac{1}{\pi}\int_{0}^{\pi} (B(\Delta\varphi)-V_{\rm Bass})d(\Delta\varphi)\approx 10\MeV$. 

Another possibility is to investigate asymmetric reactions like $^{90-96}$Zr+$^{124}$Sn. 
Despite the fact that the extra-push model predicts that the extra-push energy
becomes smaller with increasing the neutron 
excess, the experimental data suggest the opposite trend~\cite{liang2012}. 
TDHF calculations also show similar disagreement~\cite{Washiyama(2015)}.
The measured trend is consistent with the results presented here, as the 
fusion reaction is hindered as the system departs from the neutron magic $^{90}$Zr. 
The chemical potentials for colliding nuclei are fairly similar admitting
the description within broken particle-number symmetry.
We have performed exploratory simulations for asymmetric reactions, and we have
found that, similarly to the symmetric case, the phase difference can 
hinder the fusion for energies around the barrier, however no clear
solitonic structure was observed~\cite{SM}.

Finally, we have also performed simulations of non-central collisions.
If we are in the energy window where the phase difference can hinder
the fusion, we find that it affects the contact time, and consequently
the scattering angle is affected (see \cite{SM} for movies demonstrating this effect).

\ssec{Summary}
We have investigated collisions of medium and heavy nuclei at energies
around the Coulomb barrier taking into account the pairing field dynamics
with TDDFT for superfluid systems. We have found that during collision
a stable soliton-like structure appears when two superfluid nuclei collide
with phase difference of the pairing fields close to $\Delta{\varphi}=\pi$.
The solitonic excitation suppresses the neck formation and hinders energy
dissipation as well as fusion reaction, leading to significant changes in
reaction dynamics. It implies that the pairing field dynamics effectively
increases the barrier height for fusion resembling ``thud wall'' in the
extra-push model, although at much smaller energies. The Josephson current
between two colliding nuclei turns out to be small, does not exceed $2$ particles. 
The effects on the kinetic energy of the fragments and fusion cross section
may likely be observed experimentally.
Last but not least, it is to be reminded that the effects studied in this Letter are clearly
beyond the commonly used TDHF+BCS approach~\cite{Mainlacroix,Mainebata,Scamps(2013), Scamps(2015), Ebata(2015)} (see \cite{SM} for a detailed discussion).
 
\begin{acknowledgments}
We are in particular grateful to George Bertsch and Aurel Bulgac 
for discussions and critical remarks. We would like also to thank 
Nicholas Keeley, Michal Kowal, Eryk Piasecki, Krzysztof Rusek,
Janusz Skalski for helpful discussions. We thank Witold Rudnicki,
Franciszek Rakowski, Maciej Marchwiany and Kajetan Dutka from the
Interdisciplinary Centre for Mathematical and Computational Modelling (ICM)
of Warsaw University for useful discussions concerning the code optimization.
This work was supported by the Polish National Science Center (NCN) under
Contracts No. UMO-2013/08/A/ST3/00708. The code used for generation of
initial states was developed under grant of Polish NCN under Contracts 
No. UMO-2014/13/D/ST3/01940. Calculations have been performed at HA-PACS
(PACS-VIII) system---resources provided by Interdisciplinary Computational
Science Program in Center for Computational Sciences, University of Tsukuba.
The contribution of each one of the authors has been significant and
the order of the names is alphabetical.
\end{acknowledgments}

\ssec{Note added} Recently, in a related work~\cite{arXiv:1701.06683}, it
has been shown that the phase difference can influence the
outcome of the collision only in the case of systems
characterized by the weak pairing correlations (like systems
discussed here). In the strong pairing limit, the role of the
initial phase difference is erased.


\newpage
\begin{center}
{\bf Supplemental online material for:}\\
{\bf ``Novel Role of Superfluidity in Low-Energy Nuclear Reactions''}\\
\end{center}

\begin{small}
\noindent
In this supplemental material, we describe technical aspects related to: 
generation of the initial configurations, setting initial conditions for
a collision and extracting kinetic energy of the fragments after collision.
We also provide the evaluation of the contact energy of two superfluids within 
the Ginzburg-Landau theory and the classical fusion cross section taking into
account nonzero phase differences between nuclear pairing fields. We show
results of $^{240}$Pu+$^{240}$Pu with $\Delta\varphi_p \neq \Delta\varphi_n$
quantifying contributions from neutrons and protons, and with ternary
quasifission processes. We also present typical results for asymmetric
collisions of $^{86}$Zr+$^{126}$Sn and non-central collisions of $^{90}$Zr+$^{90}$Zr.
Finally, the difference between TDHF+BCS and TDHFB approaches is clarified.
Description of various supplemental movies is given.
\end{small}

\section{TDSLDA calculations for nuclear reactions}

In this section we present the methodology which has been applied in order to:
\begin{itemize}
\item prepare initial configurations with two spatially separated nuclei, 
\item imprint the phase difference of the pairing fields of the two nuclei, 
\item collide them to simulate nuclear reactions within the framework of TDSLDA.
\end{itemize}

The initial states for the TDSLDA calculations were obtained as 
self-consistent solutions of the static SLDA equations which for 
both protons and neutrons have the following structure:
\begin{equation}
\begin{pmatrix}
h-\mu    & \Delta \\
\Delta^* & -(h^*-\mu)
\end{pmatrix}
\begin{pmatrix}
u_k \\
v_k
\end{pmatrix}
=
\varepsilon_k
\begin{pmatrix}
u_k \\
v_k
\end{pmatrix},\label{eqn:hamiltonianSupp}
\end{equation}
where $h$ is the single-particle Hamiltonian and $\Delta$ is the 
pairing field, which are defined by functional derivatives of an
energy density functional, and $\mu$ is the chemical potential (for
either protons or neutrons). To reduce the number of diagonalizations
needed to get the self-consistent solution, we have used the procedure 
similar to the one adopted to compute initial states for a vortex-nucleus
system in the neutron star crust~\cite{VortexAAA}. The lattice size is
$64\times 20\times 20$ for head-on collisions and $64\times 48\times 20$
for non-central collisions. The lattice spacing is $1.25\fm$.
The pairing part of the density functional has a local form,
$\mathcal{E}_{\text{pair}}(\bm{r}) = g[|\nu_n(\bm{r})|^2 + |\nu_p(\bm{r})|^2]$, where $\nu_{p,n}$
are proton and neutron anomalous densities and the coupling constant was set to $g=-200\;\textrm{MeV fm}^3$~\cite{SuppBulgacPRL200AAA}. It corresponds to the zero-range paring force which needs to be regularized. We use the procedure described in Refs.~\cite{SuppPRL__2002AAA,SuppPRC__2002AAA}. After the regularization the pairing filed is given by $\Delta_{n,p}(\bm{r})=-g_{\textrm{eff}}\,\nu_{n,p}(\bm{r})$, where $g_{\textrm{eff}}$ is the effective coupling constant and $\nu_{n,p}$ is computed in the restricted quasiparticle space
with cutoff energy, $E_{\rm cut}=100\MeV$.

\begin{figure}[t]
   \begin{center}
   \includegraphics[width=6cm]{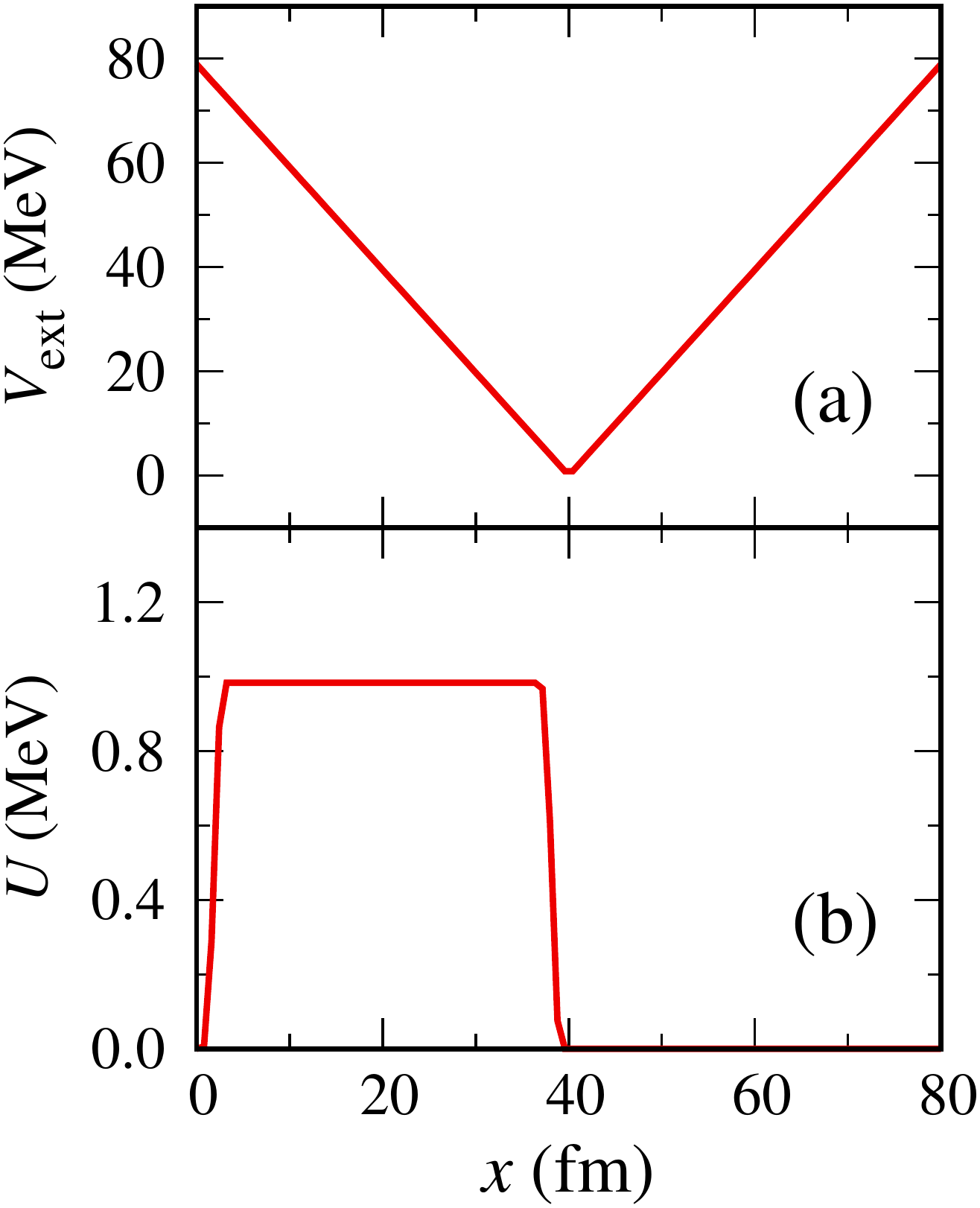}
   \end{center}\vspace{-3mm}
   \caption{
   An example of external potentials used for keeping nuclei
   at rest or for accelerating them (a) and for imprinting
   the required phase difference (b) as a function of $x$ coordinate
   (along the collision axis). (a) $V_{\rm ext}$ defined by
   Eq.~(\ref{eq:V_ext}) is shown for $V_0/\hbar c=10^{-4}$~fm$^{-1}$ and
   $\delta/\hbar c=0.0003$~fm$^{-1}$. (b) $U$ defined by Eq.~(\ref{eq:U_ext}) is
   shown for the case of imprinting the phase difference
   $\Delta\varphi=\pi$ during the time interval $1000$~fm/$c$.
   }
   \label{fig:V_ext}
\end{figure}

One has to keep in mind that $u$-components and $v$-components,
forming quasiparticle wave functions, behave differently when
expressed in the coordinate representation. While the $v$-components
are spatially localized around the two nuclei (as is always true for
bound systems), the $u$-components are distributed over the whole space.
Thus, one has to pay a particular attention when dealing with a system
of two spatially separated nuclei, as they are entangled through the
common $u$-components. For example, a discontinuity of the $u$-components
is introduced if one generates separately ground states for two nuclei
placed in smaller volumes and then combines them together in a larger volume. 
This is a typical method used in TDHF calculations, which is justified since 
the $u$- and $v$-components are decoupled for $\Delta=0$ and one evolves
only localized single-particle wave functions. In our case, however,
this method is not justified and therefore we generated self-consistent
solutions for two nuclei within a single box separated by a desired distance
$\Delta x\approx 50\fm$. In order to avoid two nuclei to move apart
because of the Coulomb repulsion, we have introduced an external potential:
\begin{equation}
V_{\rm ext}(\boldsymbol{r}) = \sqrt{[V_0 (x-x_0)]^2 + \delta^2},
\label{eq:V_ext}
\end{equation}
which is uniform in $y$- and $z$-direction. The parameter $\delta$ is
a small constant which makes $V_{\rm ext}$ smooth around the center of
the box ($x_0=40$~fm). Away from the center of the box the external
potential is linear $V_{\rm ext}(\boldsymbol{r})\simeq V_0|x-x_0|$,
and generates the constant force which compensates for the Coulomb
repulsion. The parameter $V_0$ is adjusted to keep the two nuclei at
rest during the self-consistent iterations. An example of the potential
$V_{\rm ext}$ is shown in Fig.~\ref{fig:V_ext}~(a). We used a shifted
conjugate orthogonal conjugate gradient (COCG) method to compute densities
during iterations~\cite{COCGSuppAAA}. Subsequently a direct diagonalization
of the Hamiltonian~(\ref{eqn:hamiltonianSupp}) was performed, which provided
the wave functions determining the initial configuration for both the projectile
and the target nuclei contained in the common simulation box.

The generated initial states are characterized by the pairing field that
has the uniform phase over the box. One can change the phase of
one of the nuclei without affecting the energy of the system. This can be
done dynamically, using the phase imprint technique commonly used in
experiments on ultracold atomic gases. Namely, the additional external
potential is applied for a certain time interval $t_p$. The external
potential has the following form:
\begin{equation}
U(\boldsymbol{r})=\left\lbrace 
\begin{array}{ll}
   U_0\,s(x,3.75,2),       & x \leqslant 3.75,       \\
   U_0,                    & 3.75 < x < 36.25,       \\
   U_0\,s(x-36.25,3.75,2), & 36.25 \leqslant x < 40, \\
   0,                      & x\geqslant 40,
\end{array}
\right.
\label{eq:U_ext}
\end{equation}
where $s(x,w,\alpha)$ is a function which smoothly varies from $0$ to $1$ in an interval $[0,w]$:
\begin{equation}
 s(x,w,\alpha)=\dfrac{1}{2}+
 \dfrac{1}{2}\tanh\left[ \alpha\tan\left(  \frac{\pi x}{w}-\frac{\pi}{2} \right) \right].
\end{equation}
An example of the potential is shown in Fig.~\ref{fig:V_ext}~(b).
Since the pairing field is proportional to the anomalous density 
$\nu=\sum_{0<E_n<E_{\rm cut}}u_{n} v_{n}^*$, 
the phase of the pairing field for the left half of the box ($x<40$~fm)
evolves in time as $\Delta(\boldsymbol{r},t) = e^{2i(\mu - U_0)t/\hbar}|\Delta(\boldsymbol{r},t)|$,
whereas for the right half ($x>40$~fm) it evolves as $\Delta(\boldsymbol{r},t)
= e^{2i\mu t/\hbar}|\Delta(\boldsymbol{r},t)|$. Consequently after time $t_p$
the phase for one side gets an extra shift $\Delta\varphi=2U_0t_p/\hbar$.
The height of the potential $U_0$ is adjusted to introduce the requested phase
difference $\Delta\varphi$ within a time interval $t_p=1000$~fm/$c$.

Finally, to collide two nuclei one needs to accelerate them up to a certain
value of relative velocity. It is achieved by varying smoothly the slope parameter
$V_0$ in Eq.~(\ref{eq:V_ext}) to a larger value $V_1(>V_0)$ using the switching
function. It is performed in a relatively short time of about $10\fm/c$. 
Subsequently, the slope parameter $V_1$ is kept fixed until the two nuclei
reach the desired relative velocity. Once this velocity is reached the external
potential~(\ref{eq:V_ext}) is switched off (within the time of 10~fm/$c$), and
the two nuclei collide. The collision energy is defined at the time when
$V_{\rm ext}$ becomes zero.

\section{Energy of the junction}

\begin{figure}[t]
   \begin{center}
   \includegraphics[width=7cm]{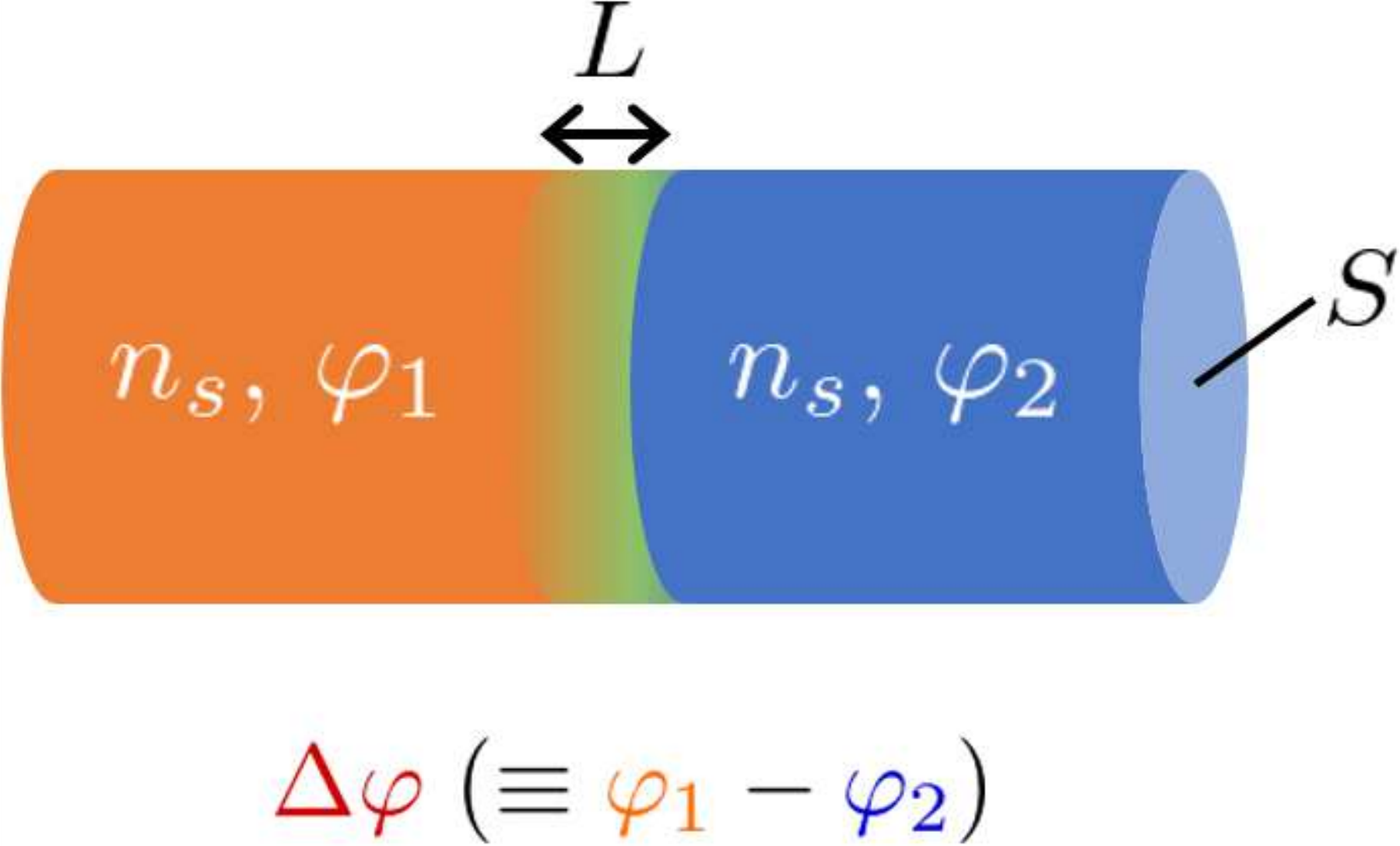}
   \end{center}\vspace{-3mm}
   \caption{
   Schematic picture of a junction of two superfluids
   with different phases. This picture illustrates those
   quantities which enters Eq.~(\ref{Eq:junction}):
   $S$ is the area of the junction, $L$ is the
   length scale over which the phase varies, $n_s$
   is the superfluid density, and $\varphi_i$ ($i=1,2$)
   is the phase of the pairing field.
   }
   \label{fig:junction}
\end{figure}

Let us consider two superfluids being in contact and having the
same superfluid density $n_s$, differing only by the phase of the
order parameter $\varphi_{1,2}$ (see Fig.~\ref{fig:junction}). 

The energy of the junction is associated with the spatial variations
of the phase on the length scale defined by the contact size. It can
be easily evaluated within the Ginzburg-Landau (G-L) approach. This is 
a macroscopic theory that describes evolution of the order parameter
of a superfluid. The only requirement is that the system is characterized
by a complex order parameter. Namely,
one can introduce the wave function of the condensate 
$\Psi(\boldsymbol{r})$ which is related to the superfluid order parameter
with $|\Psi|^2$ being the density of particles that belong to the condensate.
This quantity will be temperature dependent and clearly vanishes at
critical temperature $T=T_{c}$. Superfluidity in fermionic systems is
regarded as a condensate of Cooper pairs, thus $2|\Psi|^2$ has meaning
of fermionic density. Consequently, the wave function is expressed as
\begin{equation}
\Psi(\boldsymbol{r}) = \sqrt{\frac{n_s(\boldsymbol{r})}{2}}\,e^{i\varphi(\boldsymbol{r})},
\end{equation}
where $n_s\propto (1-\frac{T}{T_{c}})$ is the superfluid density of
fermions, and $\varphi$ denotes the phase. In general the total 
density of particles separates into superfluid and normal components
$n=n_s+n_n$ and in the zero temperature limit the latter vanishes.
The total free energy of the system is given by
\begin{equation}
F_{\rm tot} = \int
\bigl( F_L(\boldsymbol{r}) + F_{\rm grad}(\boldsymbol{r}) \bigr)\,d\boldsymbol{r},
\end{equation}
where $F_L = \alpha(T) |\Psi(\boldsymbol{r})|^2 + \beta(T) |\Psi(\boldsymbol{r})|^4 $
is the so-called Landau term. Information about type of system is encapsulated in coupling constants appearing in this term. 
Since superfluid densities of both
superfluids are equal this term does not contribute to the free
energy of the junction. What matters is the gradient term which has the form:
\begin{equation}
F_{\rm grad} = \frac{\hbar^2}{2m^*} |\nabla\Psi(\boldsymbol{r})|^2,
\end{equation}
where $m^*$ is the effective mass of particles that form the condensate,
and we assume $m^*=2m_n$ which is the mass of the Cooper pair ($m_n$
is the nucleon mass). Approximating the gradient by
\begin{equation}
\nabla\Psi(\boldsymbol{r})
\approx \sqrt{\frac{n_s}{2}}\,\frac{e^{i\varphi_2}-e^{i\varphi_1}}{L},
\end{equation}
where $L$ is the length scale over which the phase varies from value
$\varphi_1$ to $\varphi_2$, one finds the free energy of the junction
(see also Fig.~\ref{fig:junction}):
\begin{equation}
F_{j}=\frac{S}{L}\frac{\hbar^{2}}{2m_n}n_{s}\sin^{2}\frac{\Delta\varphi}{2},
\label{Eq:freejunction}
\end{equation}
where we have applied $|e^{i\varphi_2}-e^{i\varphi_1}|^2
= 4\sin^2\frac{\Delta\varphi}{2}$, denoting $\Delta\varphi = \varphi_{1} - \varphi_{2}$ 
and $SL$ is the volume of the junction. In the zero temperature limit
the free energy becomes the energy of the junction and the superfluid
density $n_{s}$ becomes neutrons/protons density:
\begin{equation}
E_{j}=\frac{S}{L}\frac{\hbar^{2}}{2m_n}n_{s}\sin^{2}\frac{\Delta\varphi}{2}.
\label{Eq:junction}
\end{equation}
For a collision of two heavy
nuclei at energies close to the Coulomb barrier, one may assume the
area of the junction $S$ corresponds to the neck, which is on the order
of $\pi R^{2}$ with $R\approx6$~fm, and taking $L\approx R$ and $n_s$
as a half of nuclear density, one finds that the energy of the junction
varies by several tens of MeV

\section{Phase evolution in a colliding system}

\begin{figure}[t]
   \begin{center}
   \includegraphics[width=\columnwidth]{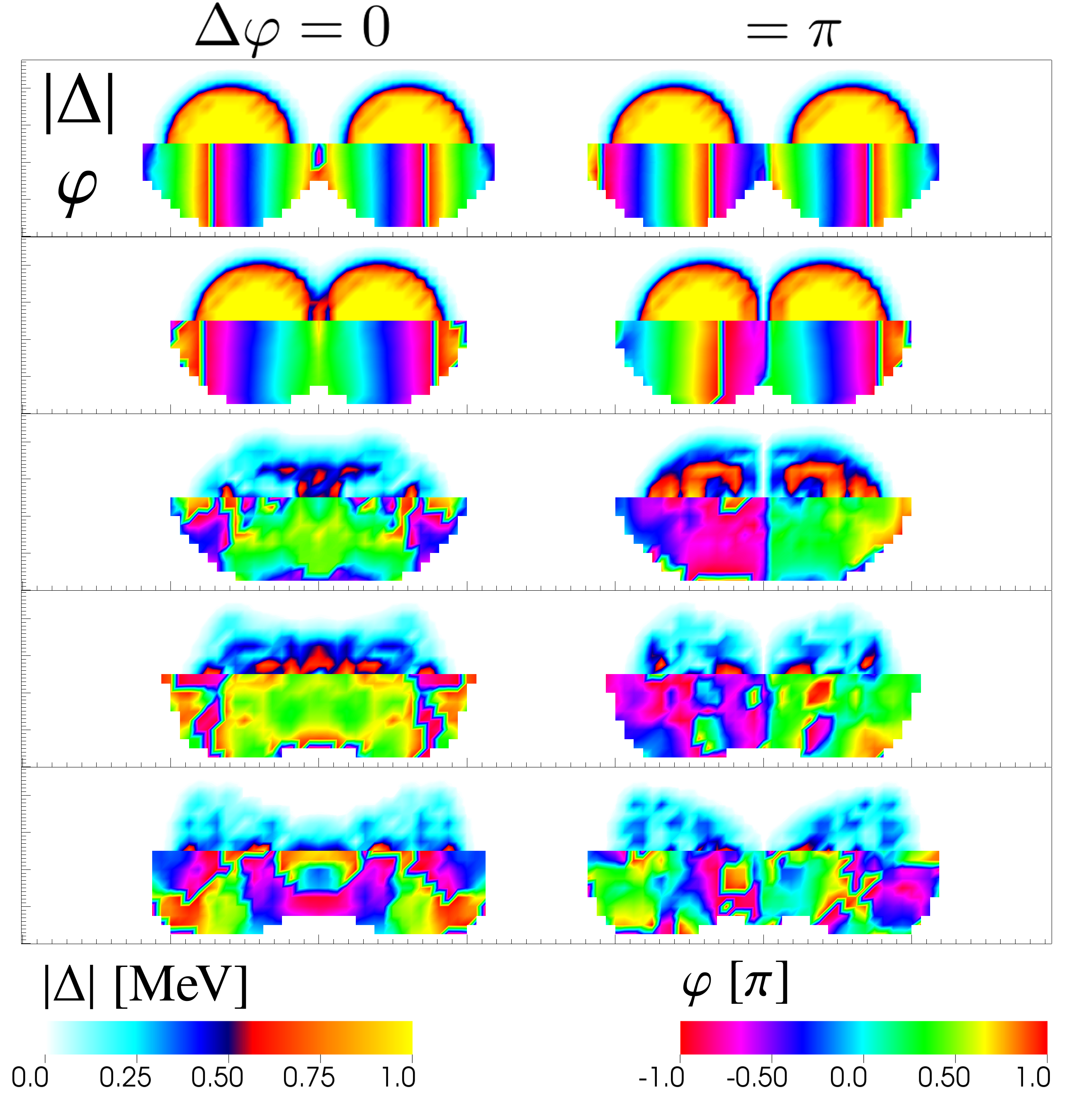}
   \end{center}\vspace{-3mm}
   \caption{
Evolution of the neutron paring field $\Delta(\boldsymbol{r})=|\Delta(\boldsymbol{r})|e^{i\varphi(\boldsymbol{r})}$ for the $^{240}$Pu+$^{240}$Pu reaction at $E \simeq 1.1V_{\rm Bass}$. Upper-half of each panel shows absolute value $|\Delta(\boldsymbol{r})|$, while lower-half presents the phase $\varphi(\boldsymbol{r})$. Left column corresponds to the case without relative phase difference between colliding nuclei in the initial state ($\Delta\varphi=0$), while right column corresponds to the extreme case where the phase difference is the largest ($\Delta\varphi=\pi$).
   }
   \label{fig:PuPuphase}
\end{figure}

In this section we provide information about time evolution of the phase of the paring field during a collision. In \fig{fig:PuPuphase}, we show the phase evolution for the $^{240}$Pu+$^{240}$Pu reaction at $E \simeq 1.1V_{\rm Bass}$, as a typical example. These snapshots correspond precisely to the frames shown in Fig.~2 of the main text. Two extreme cases corresponding to the relative phase $\Delta\varphi=0$ (left column) and $\pi$ (right column) are shown. Note that these values specify the phase difference for the initial state, where two nuclei are far away from each other. The uniformity of the phase across the nuclei is destroyed by the accelerating potential, see Eq.~(\ref{eq:U_ext}) and Fig.~\ref{fig:V_ext}\,(a).  For this reason before collision the phase varies along the collision axis inside each nucleus, see top two panels of \fig{fig:PuPuphase}. The phase relaxes fast, and when two nuclei collide and merge, one can easily define domains with well defined phase, as seen on middle panels of the figure. Finally, when the composite system splits into two fragments, the phase pattern starts to be disordered. One should note, however, that due to large excitations in this violent collision, the neutron paring is also substantially destroyed (cf. upper-half of each panel of the figure).

\section{Kinetic energy of the fragments}

The total kinetic energy (TKE) of the outgoing fragments is evaluated
as follows. We divide the simulation box into two regions by a plane
parallel to the $yz$-plane, which defines left ($x<40$~fm) and right
($x>40$~fm) regions. Subsequently we compute the average mass and charge
numbers and the center-of-mass of the fragments in respective regions, $A_{L,R}$,
$Z_{L,R}$ and $\boldsymbol{R}_{L,R}(t)$. From the time derivative of the
relative distance $\boldsymbol{R}(t) =\boldsymbol{R}_R(t)-\boldsymbol{R}_L(t)$,
we compute the relative velocity $\boldsymbol{V}(t)$. We compute the TKE
of the fragments when they are well separated spatially ($R \simeq 30$~fm)
as follows:
\begin{equation}
{\rm TKE}
= \frac{1}{2}\mu(t)\boldsymbol{V}^2(t)
+ \frac{Z_L(t)Z_R(t)e^2}{|\boldsymbol{R}(t)|},
\label{eq:TKE1}
\end{equation}
where $\mu(t)=m_n A_L(t)A_R(t)/(A_L(t)+A_R(t))$ with $m_n$ being the nucleon
mass. To check the validity of this simple formula (\ref{eq:TKE1}),
we have also computed the TKE as follows:
\begin{equation}
{\rm TKE}
= \frac{\boldsymbol{P}_L^2}{2M_L(t)}+\frac{\boldsymbol{P}_R^2}{2M_R(t)} +V_{\rm Coul}(t),
\label{eq:TKE2}
\end{equation}
where $M_{L(R)}(t)=m_n A_{L(R)}(t)$, and the momentum is defined by
\begin{equation}
\boldsymbol{P}_{L(R)}=
m_n \int_{L(R)} \boldsymbol{j}(\boldsymbol{r},t)\,d\boldsymbol{r}.
\end{equation}
The Coulomb energy can be evaluated from the proton density:
\begin{equation}
V_{\rm Coul}(t) = e^2 \int_L\int_R
\frac{\rho_p(\boldsymbol{r}_1,t)\rho_p(\boldsymbol{r}_2,t)}{|\boldsymbol{r}_1-\boldsymbol{r}_2|}\,
d\boldsymbol{r}_1d\boldsymbol{r}_2.
\end{equation}
We have found that the difference between Eqs.~(\ref{eq:TKE1}) and (\ref{eq:TKE2})
is very small, at most a few MeV at certain times where fragments are close
to each other and exhibit large deformations, where multipole corrections
play a role.

\section{Effects on fusion cross section}

The effect of the increased barrier can be included in the expression for the
classical fusion cross section. Namely, the classical expression reads:
\begin{equation}\label{classical}
\sigma(E) = \pi R^{2} \left ( 1-\frac{B}{E} \right )
\end{equation}
where $B$ defines the barrier height and $E>B$. The modification of this expression
related to the pairing field phase difference reads:
\begin{equation}
\sigma(E)= R^2 \left (\Delta\varphi_{\rm th}(E) - \frac{1}{E}\int_{0}^{\Delta\varphi_{\rm th} (E)} B(\Delta\varphi)d(\Delta\varphi) \right ),
\end{equation}
where $R = r_{0}(A_{1}^{1/3} + A_{2}^{1/3})$ with $r_{0}\approx1.25$~fm,
and $\Delta\varphi_{\rm th}$ is the threshold phase difference below which the capture occurs
at a given collision energy $E$. The expression requires $E \ge E_{\rm min}$,
where $E_{\rm min}$ is the lowest energy for capture without phase difference,
\textit{i.e.}, $\Delta\varphi_{\rm th}(E_{\rm min})=0$. One can easily notice that
for $E>E_{\rm max}$, where $\Delta\varphi_{\rm th}(E_{\rm max})=\pi$, the above
formula is consistent with Eq.~(\ref{classical}), with barrier height averaged
over all angles: $\bar{B}=\frac{1}{\pi}\int_{0}^{\pi} B(\Delta\varphi)d(\Delta\varphi)$.
Thus the effect of the phase difference on the fusion cross section will enter through
the effective barrier height averaged over all phase differences. In the energy
interval $E_{\rm min} < E < E_{\rm max}$ one may use the above formula to extract
energy dependence of $\Delta\varphi_{\rm th}$, directly from the excitation function, namely,
\begin{equation}
\frac{1}{R^{2}}\frac{d}{dE}(E\sigma(E))= \Delta\varphi_{\rm th}(E).
\end{equation}
In practice, however, this quantity will be contaminated by other effects like,
\textit{e.g.}, quantum tunneling and nonzero width of the barrier distribution.

\begin{figure}[t]
   \begin{center}
   \includegraphics[width=7cm]{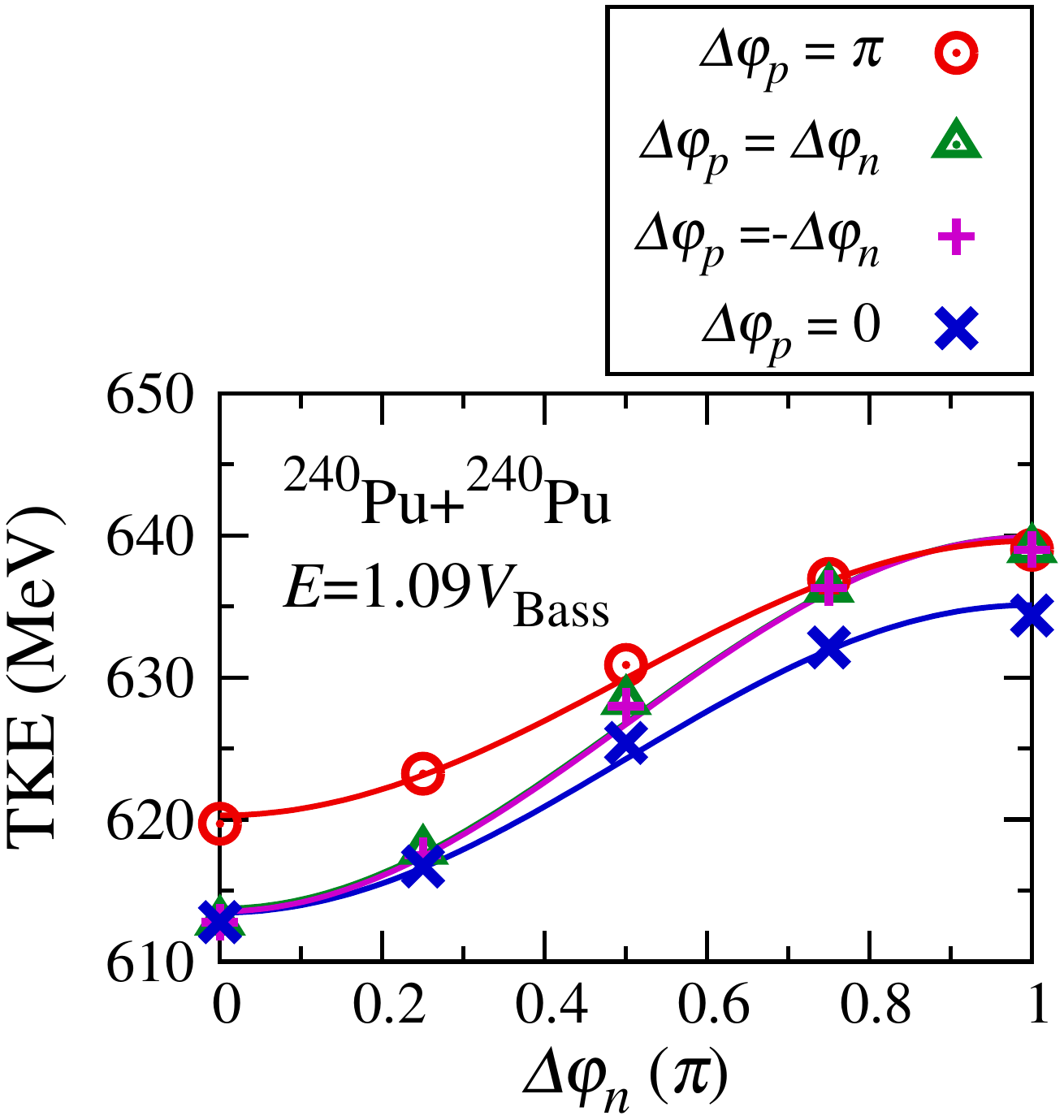}
   \end{center}\vspace{-3mm}
   \caption{
   Total kinetic energy (TKE) of outgoing fragments in
   $^{240}$Pu+$^{240}$Pu collisions at $E\simeq1.09V_{\rm Bass}$
   as a function of the pairing phase difference for neutron superfluids, $\Delta\varphi_n$.
   The four different symbols correspond to the following cases:
   green open triangles denote results obtained under condition that the imprinted phase 
   difference for both neutron and proton superfluids is the same: $\Delta\varphi_p=\Delta\varphi_n=
   \Delta\varphi$ (This case has been shown in Fig.~3~(a) of the main text);
   pink plus symbols correspond to the case of proton superfluids having the opposite phase difference
   to that of neutron superfluids: $\Delta\varphi_p=-\Delta\varphi_n$;
   red open circles (blue crosses) correspond to the case of proton superfluids
   having the fixed phase difference at values: $\Delta\varphi_p=\pi$ ($0$).
   The solid curves represent fits to the data points with the expression
   $\alpha+\beta\sin^2\frac{\Delta\varphi_n}{2}$, taking $\alpha$ and $\beta$ as parameters.
   }
   \label{fig:Pu+Pu_TKE}
\end{figure}

\begin{figure*}[t]
   \begin{center}
   \includegraphics[width=0.96\textwidth]{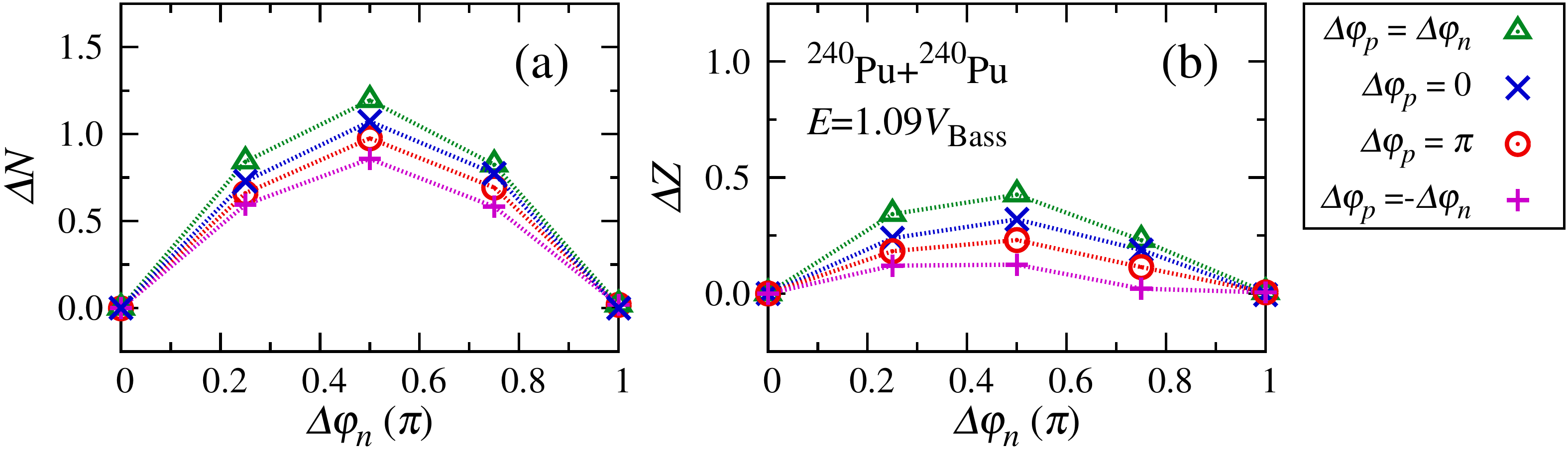}
   \end{center}\vspace{-3mm}
   \caption{
   Average number of transferred neutrons (a) and protons (b)
   from the left to the right regions in $^{240}$Pu+$^{240}$Pu collisions
   at $E \simeq 1.09V_{\rm Bass}$ for various types of phase
   differences. The horizontal axis is the pairing phase difference
   for neutron superfluids, $\Delta\varphi_n$.
   Green open triangles show results where the phase is imprinted 
   for both neutrons and protons the same amount ($\Delta\varphi_p=\Delta\varphi_n=
   \Delta\varphi$) (The same results as shown in Fig.~3~(b) of the main text).
   Pink plus symbols correspond to the cases where the phase difference
   for protons is opposite to that for neutrons ($\Delta\varphi_p=-\Delta\varphi_n$).
   Red open circles (blue crosses) correspond to the cases where the phase 
   difference for protons is fixed to $\Delta\varphi_p=\pi$ ($0$).
   }
   \label{fig:Pu+Pu_dN}
\end{figure*}

\section{The case of $\Delta\varphi_p \neq \Delta\varphi_n$}

In order to investigate the magnitude of contributions 
coming from proton and neutron superfluids separately,
we have performed simulations of head-on collisions of $^{240}$Pu+$^{240}$Pu
with different relative phases of proton and neutron pairing fields, 
\textit{i.e.}~$\Delta\varphi_p \neq \Delta\varphi_n$.

Let us first focus on the TKE of the fragments.
In Fig.~\ref{fig:Pu+Pu_TKE}, we show the TKE as a function of the
pairing phase difference for neutron superfluids, $\Delta\varphi_n$. The collision
energy was set to $E \simeq 1.09V_{\rm Bass}$, at which we have observed 
the most pronounced effect generating the largest TKE differences of 
about $25$~MeV (see Fig.~3~(a) in the main text). The quantity $V_{\rm Bass}$ is the
phenomenological fusion barrier~\cite{SupBass1974AAA}.

In the figure one can see four curves representing TKE($\Delta\varphi_n$),
although two of them ($\Delta\varphi_p=\pm\Delta\varphi_n$ cases)
practically coincide. It indicates that the effect governing the 
TKE behavior is related to the solitonic excitation, as the energy of the junction is
the same in the two cases. This fact also confirms that the effects related 
to Josephson currents are indeed negligible, since in the case of $\Delta\varphi_p=
\Delta\varphi_n$ protons and neutrons are transferred in the
same direction, whereas in the case of $\Delta\varphi_p=-\Delta\varphi_n$
the directions of induced currents for protons and neutrons are opposite.
The other two curves correspond to $\Delta\varphi_p=\pi$ and $\Delta\varphi_p=0$
cases. The relative energy shift of these curves measures the magnitude of 
the contribution coming from the pairing phase difference for protons. Thus, clearly
the total effect reflected in TKE comes from both proton and neutron superfluids,
however the neutron contribution ($\approx 21$~MeV) is significantly larger than
the proton contribution ($\approx 4\mbox{--}7$~MeV). This we attribute to the 
fact that neutrons play more important role in the neck formation, whereas
contribution of protons is effectively suppressed by the Coulomb repulsion.

In Fig.~\ref{fig:Pu+Pu_dN}, we show the average number of
transferred neutrons (a) and protons (b) from the left nucleus 
to the right one as a function of the pairing phase difference for
neutrons, $\Delta\varphi_n$. All symbols are the same as in
Fig.~\ref{fig:Pu+Pu_TKE}. In particular, green open triangles shown
in Fig.~\ref{fig:Pu+Pu_dN}~(a) denote the results shown also in
Fig.~3~(b) of the main text. The amount of nucleons transferred
during the collisions is the largest for
the case of $\Delta\varphi_p=\Delta\varphi_n$, \textit{i.e.}, the case
where both protons and neutrons flow in the same direction.
On the other hand, in the case of $\Delta\varphi_p=-\Delta\varphi_n$ 
the amount of transferred nucleons is the smallest, since induced 
currents for protons and neutrons have the opposite directions, and due to the
mutual entrainment the magnitude of the currents is suppressed. Moreover,
one can find a visible difference between the cases of $\Delta\varphi_p=0$
and $\Delta\varphi_p=\pi$, despite the fact that the proton induced current 
is absent in both cases. It indicates that the solitonic structure in the 
proton pairing field plays also a role of a barrier suppressing neutron flow.
Nevertheless in both cases ($\Delta\varphi_p=0,\pi$) a small number of protons 
are transferred, it is only due to the neutron induced current which entrain protons.

\section{Asymmetric reactions}

\begin{figure}[t]
   \begin{center}\vspace{1mm}
   \includegraphics[width=7cm]{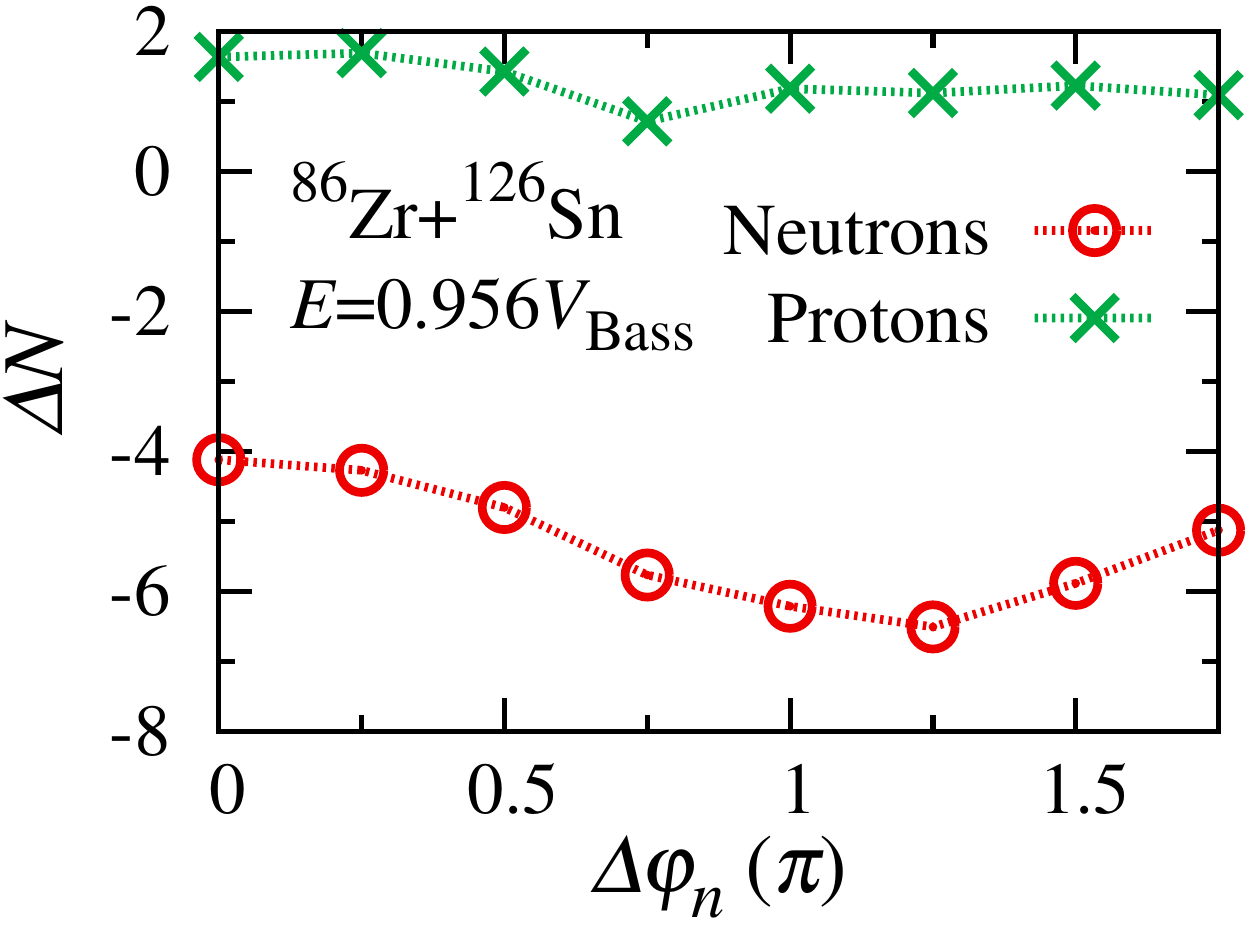}
   \end{center}\vspace{-3mm}
   \caption{
   Average number of transferred neutrons (red circles) and protons (green crosses) in an asymmetric reaction, $^{86}$Zr+$^{126}$Sn, at an energy below the fusion threshold ($E=0.956 V_{\rm Bass}$) as a function of phase differences for the neutron paring field, $\Delta\varphi_n$, in the initial state.
   }
   \label{fig:Zr+Sn_dN}
\end{figure}

To examine the effects of the pairing phase difference on the dynamics
in asymmetric reactions, we performed exploratory calculations for
$^{86}$Zr+$^{126}$Sn. In these reactions the phase difference is not 
a well-defined quantity, but it evolves in time and the evolution rate
is determined by the difference of the nuclear chemical potentials.
In the case of symmetric reactions (like $^{90}$Zr+$^{90}$Zr) the
collisions for the phase differences $\Delta\varphi$ and $2\pi-\Delta\varphi$,
are connected by reflection symmetry (\textit{i.e.}~the symmetry under
an exchange of the colliding nuclei) and therefore
the relevant range for the phase difference is limited to the interval $[0,\pi]$.
This is no longer correct for asymmetric reactions and one has to consider
all phase differences that span interval $\Delta\varphi\in [0,2\pi]$.
Nevertheless, we have observed qualitatively the same effect of the
barrier enhancement due to the pairing as in $^{90}$Zr+$^{90}$Zr reactions.

For the asymmetric reaction, at energy below the fusion threshold,
the natural process that takes place is the particle transfer (due to
difference in the chemical potentials). We observed that the phase difference
can modify the amount of nucleon transfer. As an example in Fig.~\ref{fig:Zr+Sn_dN}
we show the particle transfer for various phase differences of the initial state.
Depending on the relative phase of the pairing fields, 4--6 neutrons are
transferred. The fluctuations in the particle transfer are about 2 particles, 
thus very similar to the other cases. The fluctuation of 2 particles is 
due to the induced current that can either enhance or suppress the particle flow. 
The effects also induce indirectly variations for protons transfer
(note that $Z=40$ is a magic number without spin-orbit coupling
and protons in Zr are in normal phase).

\section{Ternary quasifission}

\begin{figure}[t]
   \begin{center}
   \includegraphics[width=0.95\columnwidth]{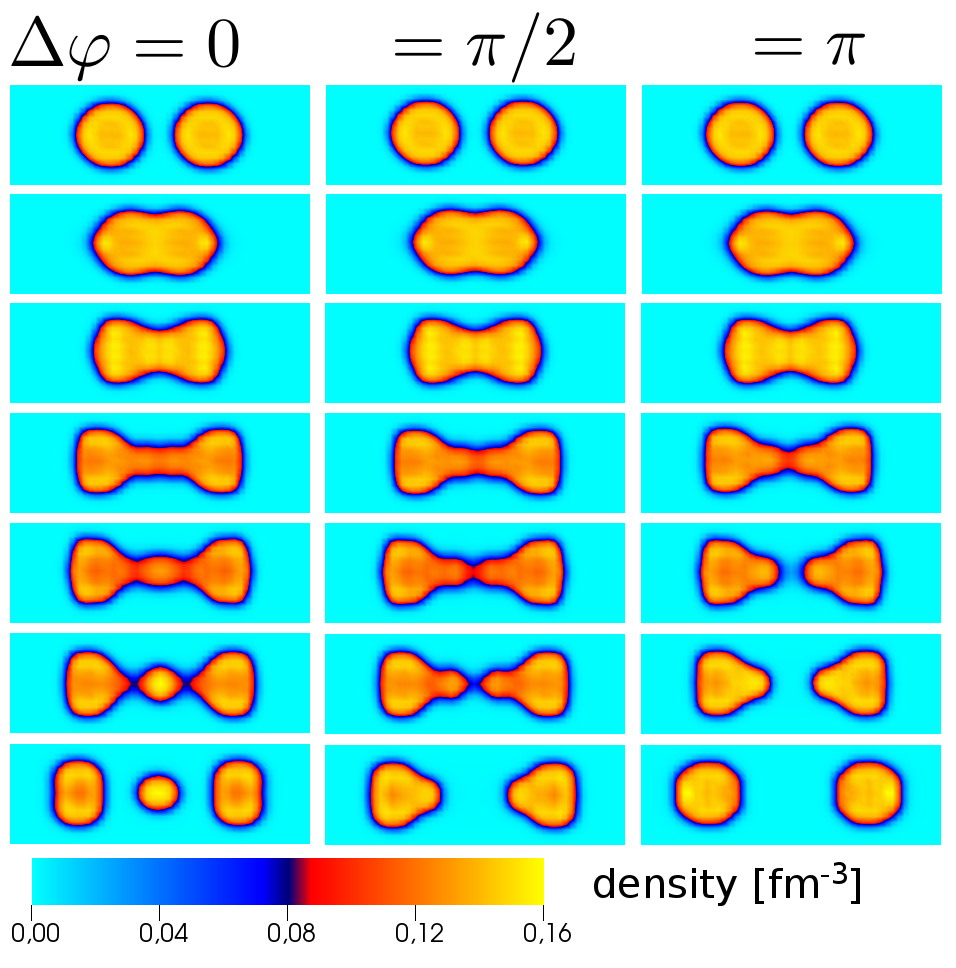}
   \end{center}\vspace{-3mm}
   \caption{
   Snapshots of the density distribution on the reaction plane for $^{240}$Pu+$^{240}$Pu collisions
   at $E \simeq 1.3V_{\rm Bass}$, for three pairing phase differences:
   $\Delta\varphi=0$ (left column), $\Delta\varphi=\pi/2$ (middle column) and $\Delta\varphi=\pi$ (right column). 
   Contact time spans the interval $570\mbox{--}650\fm/c$ depending on the phase difference. 
   A third fragment is created for the phase difference $\Delta\varphi\lesssim\frac{\pi}{4}$. 
   For full movie, see {\tt 240Pu+240Pu\_1.30V.mp4} (\url{https://youtu.be/7UstUB6DBn4}).
   }
   \label{fig:ternary}
\end{figure}

In the case of $^{240}$Pu+$^{240}$Pu collisions at sufficiently high
energies, we have observed exotic reaction dynamics. Namely, at energy
$E \simeq 1.5 V_{\rm Bass}$ the composite system splits producing a
third light fragment for all phase differences $\Delta\varphi$. Such
a ternary quasifission process has been observed as well in TDHF approach~\cite{simenelAAA},
indicating the possibility of emission of a light fragment in the collision
process. Interestingly, the third fragment is not at rest as it should
be in the case of TDHF approach, where the left-right symmetry is being
conserved in symmetric collisions in the center-of-mass frame. In our approach,
however, the symmetry is broken due to the different phases of pairing fields
of incoming nuclei. The induced current appears as a consequence
of the symmetry breaking component in the pairing field. Therefore, it is not
surprising that the third fragment is not at rest and is moving after splitting
due to the current induced by the phase difference. Moreover, at energies
$E \simeq 1.3 V_{\rm Bass}$ and $E \simeq 1.4 V_{\rm Bass}$, we have found that
the number of fragments is affected by the phase difference. The situation
is exhibited in Fig.~\ref{fig:ternary}. In this case, the solitonic structure
prevents the formation of the third fragment from the neck region.
The observed effects implicitly indicate the importance of the phase 
difference on the reaction dynamics even at relatively high energies, 
although one should keep in mind if TDSLDA description for such 
high energy collisions is valid.

\section{Non-central collisions}

\begin{figure}[t]
   \begin{center}
   \includegraphics[width=0.95\columnwidth, trim=30 80 30 80, clip]{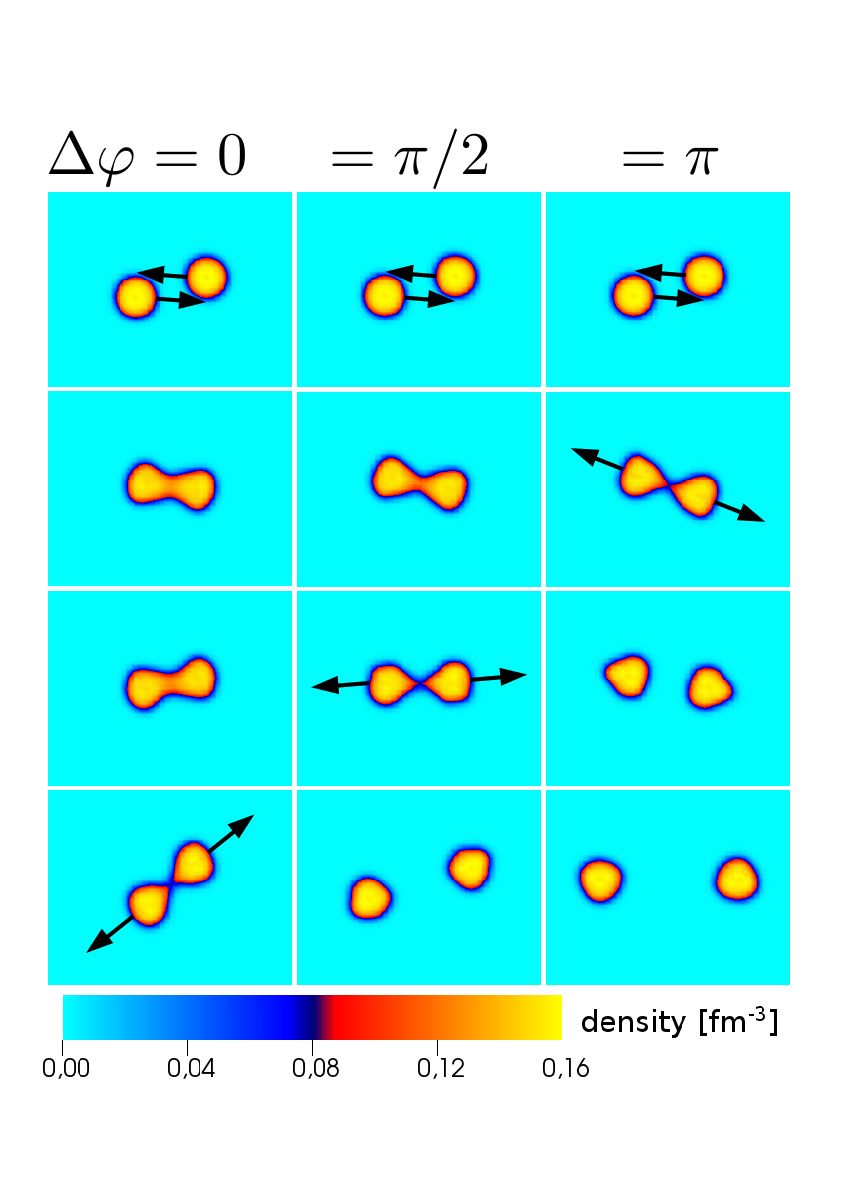}
   \end{center}\vspace{-3mm}
   \caption{
   Snapshots of the density on the reaction plane for non-central collisions of $^{90}$Zr+$^{90}$Zr 
   at $E \simeq 1.4V_{\rm Bass}$ ($b$\,$\approx$\,2~fm), for three pairing phase differences:
   $\Delta\varphi=0$ (left column), $\Delta\varphi=\pi/2$ (middle column) and $\Delta\varphi=\pi$ (right column). Contact time depends on the relative phase difference 
   and it is approximately equal to: $1000,\,840$, and $720\fm/c$, 
   for phase differences $0,\, \pi/2$, and $\pi$, respectively. 
   For full movie, see {\tt 90Zr+90Zr\_noncentral3\_1.38V.mp4} (\url{https://youtu.be/UCCAN9ahNqA}).
   }
   \label{fig:Zr_Zrcol}
\end{figure}

In order to investigate collision trajectories and their
dependence on the pairing field phase differences, we have
performed collisions with nonzero impact parameters.
Since we solve TDSLDA equations in 3D Cartesian coordinates
without any symmetry restrictions, we can also simulate
non-central collisions. We have performed simulations
for non-central collisions of $^{240}$Pu+$^{240}$Pu and 
$^{90}$Zr+$^{90}$Zr. Similarly to central collisions,
the phase difference prevents superfluid nucleons to enter
the neck region. It causes the suppression of the neck
formation and consequently the reduction of the contact time. 
In the non-central collisions, however, the small variations of
the contact time have a dramatic effect on the collision trajectories.
It affects both the scattering angle as well as kinetic
energy of the fragments and induces the correlation between these
two quantities at a fixed impact parameter. In Fig.~\ref{fig:Zr_Zrcol},
we show an example of the non-central collisions of $^{90}$Zr+$^{90}$Zr, 
where the energy and the impact parameter are chosen in a such way that
the system does not fuse, but splits after certain time. As is apparent
from the figure, the pairing phase difference affects the contact
time and consequently the scattering angles.

\section{TDHF+BCS {\em vs.} TDHFB}

In this section we demonstrate the effects that are triggered by paring field dynamics, like studied in this Letter, cannot be investigated within TDHF+BCS framework.

Let us begin with the static description. The difference
between BCS and HFB treatments can be formally expressed 
on the level of the Bloch-Messiah-Zumino 
(BMZ) decomposition of the Bogoliubov transform~\cite{BlochMessiahAAA,ZuminoAAA}. Namely, the BCS approach assumes that the third
transform of the BMZ decomposition, which mixes quasiparticle states, is equal to unity.
Physically, it means that BCS approach cannot describe processes due to the quasiparticle scattering.
Therefore the phenomena which involve interaction of quasiparticles with nonuniform pairing
field are out of range for the BCS approach. Hence, \textit{e.g.}, such phenomena, related to the nonuniformity
of the pairing field, as the existence of Andreev states or Andreev reflection, which are well-known
in the condensed matter physics (also in the context of Josephson junction) cannot be described
within the BCS framework. Since in the HF+BCS approach one finds occupation numbers of the HF orbitals, which are 
just numbers associated with each orbital, therefore it cannot describe a configuration with
position dependent phase of the pairing field. For example, the stationary vortex solution cannot be 
obtained as it requires the occupation numbers to vary when one is approaching the core of the vortex
where the system becomes normal (pairing field tends to zero).
In other words, the BCS treatment introduces only one additional degree of freedom, which is related
to the magnitude of the complex pairing field, but is unable to describe properly its excitation modes, \textit{e.g.}, Bogoliubov phonons.

The situation in the case of TDHF+BCS approach is similar. The evolution is performed through
the equations:
\begin{equation}
i\hbar\frac{\partial}{\partial t}\psi_{k}(\bm{r},t) = \hat{h} \psi_{k}(\bm{r},t),
\end{equation}
which define the evolution of HF orbitals according to the mean-field $\hat{h}$, and $|k\rangle = |\bm{k}\sigma\rangle$.
In addition, the equations describing the evolution of diagonal density matrix $\rho$ and
pairing tensor $\nu$ in this basis are solved (see Refs.~\cite{lacroixSuppAAA, ebataSuppAAA}):
\begin{eqnarray}
 \frac{d}{dt}\rho_{kk} = \Delta_{k\bar{k}}\nu_{k\bar{k}}^{*}- \Delta_{k\bar{k}}^{*}\nu_{k\bar{k}}, \\
 \frac{d}{dt}\nu_{k\bar{k}} = \Delta_{k\bar{k}}(1-2\rho_{kk}),
\end{eqnarray} 
where $\bar{k}$ is the time-reversal partner of the state $k$.
These equations describe simply the evolution of the occupation numbers $v_k$ and $u_k$ of the HF orbitals because:
\begin{eqnarray}
\rho(\bm{r},t)&=&\sum_{k}|v_k(t)|^2\,|\psi_{k}(\bm{r},t)|^2,\label{eqn:rhoBCS}\\
\nu(\bm{r},t)&=&\sum_{k}u_k(t) v_k(t)\,\psi_{k}(\bm{r},t)\psi_{\bar{k}}(\bm{r},t).
\end{eqnarray} 
The spatial dependence of these occupation numbers is not included and consequently the dynamics of the pairing 
field is realized only through changes of the HF orbitals.

In order to give a simple example of the case where the TDHF+BCS method fails completely,
let us imagine that we have a uniform system with a non-vanishing pairing gap. In such a case,
the static HF+BCS treatment is equivalent to the HFB treatment, as there 
is no quasiparticle scattering in the system and the canonical basis corresponds simply to plane waves.
Now imagine that we apply an external, spatially modulated, pairing field $\Delta_{\rm ext}(\bm{r})$. How the system will react?
In the TDHF+BCS treatment this spatial modulation cannot be described, as the system of equations
is initially in the canonical basis which corresponds to plane waves and these plane waves are eigenstates of the mean field:
\begin{equation}
\hat{h} \psi_{k}(\bm{r},t) = \varepsilon_{k} \psi_{k}(\bm{r},t).
\end{equation}
Therefore HF orbitals will not change (apart from the phase change) and they
will not be affected by the external pairing field. The modification
of the above equations after the application of the external pairing field results 
in modifying the term:
\begin{equation}
\Delta_{k\bar{k}} \rightarrow \Delta_{k\bar{k}} + \Delta^{\rm ext}_{k\bar{k}}.
\end{equation}
Clearly, $\Delta_{\rm ext}(\bm{r})$ potential enters into the above equations
in the form of the matrix element in the plane wave basis: 
$\Delta^{\rm ext}_{k\bar{k}} \propto \int \Delta_{\rm ext}(\bm{r})d\bm{r}$.
Therefore it will only change the magnitude of the pairing field on average and will result in an average 
oscillations of the uniform pairing field.
Since the density in this treatment is expressed by Eq.~(\ref{eqn:rhoBCS})
and $\psi_{k}\propto e^{i\bm{k}\cdot\bm{r}}$,
there is no mechanism to break the translational symmetry induced by the external pairing field as it only may occur through the symmetry breaking terms in the mean-field Hamiltonian (these are absent
according to our initial assumption).
This simple example clearly shows that in the TDHF+BCS treatment the degrees of freedom associated 
with the Cooper-pair dynamics are treated only in a very limited way and the pairing field dynamics is practically absent. The spatial modulation of the pairing field in the TDHF+BCS dynamics may occur only as a consequence
of the evolution of the normal density $\rho$. 

In the TDHFB treatment, on the other hand, the situation is radically different and the pairing field $\Delta(\bm{r},t)$
has its own degrees freedom treated on the same footing as the normal degrees of freedom described by 
$\rho(\bm{r},t)$. They are of course coupled but formally independent. 
Consequently, in the TDHFB evolution, the modulation of the external pairing field
will propagate leading to a variety of effects, including translational symmetry breaking
of the mean-field, and giving rise
to various quasiparticle scattering processes. These processes are beyond the simplified 
TDHF+BCS treatment. Last but not least, one has to keep in mind that TDHF+BCS equations violate the continuity equation which produce various unwanted effects (see Ref. \cite{lacroixSuppAAA}).

In summary, the effect studied in the paper is exactly of the nature that prevents its description by 
the TDHF+BCS approach. Namely, the excitation of the pairing modes in the form of a soliton induces also 
the modification of the normal density, however the dynamics is triggered by the dynamics of the pairing 
field which is induced by the spatial variation of the pairing field
and cannot be described by the TDHF+BCS approach.

\section{Movies}

\subsection{Central collisions of ${}^{240}\textrm{Pu}+{}^{240}\textrm{Pu}$}
The movies show sections along the reaction plane. Each movie displays
10 panels organized in a grid of 2 columns and 5 rows. The left column presents
the total density (density of protons + density of neutrons), while the
right column presents absolute value of the pairing field of neutrons.
In each row dynamics of the system for pairing phase difference $\Delta\varphi$
is shown. The phase differences are from $0$ (bottom row) to $\pi$ (top row)
with an increment $\pi/4$. The only difference in initial states is the phase
difference, all other quantities (like energy, density distribution, etc.)
are exactly the same (up to machine precision). Thus, all differences in
the dynamics are due to the pairing effects. Below we provide 8 movies
for different collision energies from a range $E\in [1.04\,V_{\rm Bass},1.50\,V_{\rm Bass}]$,
where $V_{\rm Bass}=897.31\MeV$ is the phenomenological fusion barrier~\cite{SupBass1974AAA}.
The value of the collision energy is encoded in the file name. 
\begin{enumerate}
\item{File: {\tt 240Pu+240Pu\_1.04V.mp4}\\
YouTube: \url{https://youtu.be/foA33kCPT5g}
}
\item{File: {\tt 240Pu+240Pu\_1.07V.mp4}\\
YouTube: \url{https://youtu.be/jiWpUUAe7Uw}
}
\item{File: {\tt 240Pu+240Pu\_1.09V.mp4}\\
YouTube: \url{https://youtu.be/0YiBJlPFVnA}
}
\item{File: {\tt 240Pu+240Pu\_1.15V.mp4}\\
YouTube: \url{https://youtu.be/ItlZQRw9yDs}
}
\item{File: {\tt 240Pu+240Pu\_1.20V.mp4}\\
YouTube: \url{https://youtu.be/bXTzRW2HgTQ}
}
\item{File: {\tt 240Pu+240Pu\_1.30V.mp4}\\
YouTube: \url{https://youtu.be/7UstUB6DBn4}
}
\item{File: {\tt 240Pu+240Pu\_1.40V.mp4}\\
YouTube: \url{https://youtu.be/rHLSWPYj798}
}
\item{File: {\tt 240Pu+240Pu\_1.50V.mp4}\\
YouTube: \url{https://youtu.be/YH0wSPoU5ag}
}
\end{enumerate}

\subsection{Central collisions of ${}^{90}\textrm{Zr}+{}^{90}\textrm{Zr}$}
The movies are analogues to the movies for ${}^{240}\textrm{Pu}+{}^{240}\textrm{Pu}$
collisions. For some energies not all panels are filled with data. These are
situations that correspond to the fusion process. Moreover, for some cases
one can notice that the system slowly rotates after collision. The effect
originates from the fact that the collisions are not perfectly central
(due to small numerical noise). Below we provide 12 movies for different
collision energies from a range $E\in [0.98\,V_{\rm Bass},1.15\,V_{\rm Bass}]$,
where $V_{\rm Bass}=192.47\MeV$ and the value of the collision energy is encoded
in the file name. 
\begin{enumerate}
\item{File: {\tt 90Zr+90Zr\_0.98V.mp4}\\
YouTube: \url{https://youtu.be/YcATJ6pMgD0}
}
\item{File: {\tt 90Zr+90Zr\_0.99V.mp4}\\
YouTube: \url{https://youtu.be/amXMsxw1Wd0}
}
\item{File: {\tt 90Zr+90Zr\_1.00V.mp4}\\
YouTube: \url{https://youtu.be/zE74gdLTgWw}
}
\item{File: {\tt 90Zr+90Zr\_1.02V.mp4}\\
YouTube: \url{https://youtu.be/UEPiwEZDeYc}
}
\item{File: {\tt 90Zr+90Zr\_1.03V.mp4}\\
YouTube: \url{https://youtu.be/Ubs3g12UW5U}
}
\item{File: {\tt 90Zr+90Zr\_1.04V.mp4}\\
YouTube: \url{https://youtu.be/GysuioTC7mY}
}
\item{File: {\tt 90Zr+90Zr\_1.05V.mp4}\\
YouTube: \url{https://youtu.be/cNbiadr7i48}
}
\item{File: {\tt 90Zr+90Zr\_1.11V.mp4}\\
YouTube: \url{https://youtu.be/7Rc7Obbnkx8}
}
\item{File: {\tt 90Zr+90Zr\_1.12V.mp4}\\
YouTube: \url{https://youtu.be/CZ_dl9vIbtI}
}
\item{File: {\tt 90Zr+90Zr\_1.13V.mp4}\\
YouTube: \url{https://youtu.be/jv5iyFFrqBI}
}
\item{File: {\tt 90Zr+90Zr\_1.14V.mp4}\\
YouTube: \url{https://youtu.be/sPH9PNEIVo4}
}
\item{File: {\tt 90Zr+90Zr\_1.15V.mp4}\\
YouTube: \url{https://youtu.be/U3t_xcdrWTA}
}
\end{enumerate}

\vspace{4mm}
\subsection{Non-central collisions of ${}^{90}\textrm{Zr}+{}^{90}\textrm{Zr}$}
We also provide 3 movies demonstrating the impact of the phase difference
on dynamics of non-central collisions. For better visibility we display
only results for 3 phase differences: $0$ (bottom row), $\frac{\pi}{2}$
(middle row) and $\pi$ (top row). The collisions are for fixed collision
energy and 3 different impact parameters. 
\begin{enumerate}
\item{File: {\tt 90Zr+90Zr\_noncentral1\_1.38V.mp4}\\
YouTube: \url{https://youtu.be/bOLhIEmfFSQ}
}
\item{File: {\tt 90Zr+90Zr\_noncentral2\_1.38V.mp4}\\
YouTube: \url{https://youtu.be/N72VQJVo4aI}
}
\item{File: {\tt 90Zr+90Zr\_noncentral3\_1.38V.mp4}\\
YouTube: \url{https://youtu.be/UCCAN9ahNqA}
}
\end{enumerate}


\begin{thebibliography}{99}
\bibitem{ring} P. Ring and P. Schuck, The Nuclear Many-Body Problem (Springer-Verlag, Berlin, 2000).
\bibitem{shimizu} Y.R. Shimizu, J.D. Garrett, R.A. Broglia, M.Gallardo, and E. Vigezzi, Pairing fluctuations in rapidly rotating nuclei,
                  Rev. Mod. Phys. {\bf 61}, 131 (1989).
\bibitem{bender} M. Bender, P-H. Heenen, and P-G. Reinhard, Self-consistent mean-field models for nuclear structure, 
                 Rev. Mod. Phys. {\bf 75}, 121 (2003).
\bibitem{dean} D.J. Dean and M. Hjorth-Jensen, Pairing in nuclear systems: from neutron stars to finite nuclei, 
               Rev. Mod. Phys. {\bf 75}, 607 (2003).
\bibitem{hashimoto2013} Y. Hashimoto, Time-dependent Hartree-Fock-Bogoliubov calculations using a Lagrange mesh with the Gogny    
                        interaction,  Phys. Rev. C {\bf 88}, 034307 (2013).
\bibitem{VortexHe} E. Fonda, D.P. Meichle, N.T. Ouellette, S. Hormoz, and D.P. Lathrop, Direct observation of Kelvin waves excited by quantized vortex reconnection, Proc. Natl. Acad. Sci. U.S.A. {\bf 111}, 4707 (2014).

\bibitem{MIT1} M.J.H. Ku, W. Ji, B. Mukherjee, E. Guardado-Sanchez, L. W. Cheuk, T. Yefsah, and M. W. Zwierlein, Motion of a Solitonic Vortex in the BEC-BCS Crossover, Phys. Rev. Lett. {\bf 113}, 065301 (2014).
\bibitem{MIT2} M.J.H. Ku, B. Mukherjee, T. Yefsah, and M.W. Zwierlein, Cascade of Solitonic Excitations in a Superfluid Fermi gas: From Planar Solitons to Vortex Rings and Lines, Phys. Rev. Lett. {\bf 116}, 045304 (2016).

\bibitem{PRL__2014} A. Bulgac, M.M. Forbes, M.M. Kelley, K.J. Roche, and G. Wlaz\l{}owski,
              Quantized Superfluid Vortex Rings in the Unitary Fermi Gas,
              Phys. Rev. Lett. {\bf 112}, 025301 (2014).  
\bibitem{PRA__2015} G. Wlaz\l{}owski, A. Bulgac, M.M. Forbes, and K.J. Roche, 
              Life Cycle of Superfluid Vortices and quantum turbulence in the Unitary Fermi Gas, 
              Phys. Rev. A {\bf 91}, 031602(R) (2015).
\bibitem{becbcs} The BCS-BEC Crossover and the Unitary Fermi Gas, Lecture Notes in Physics {\bf 836}, ed. W. Zwerger, 
                 (Springer-Heidelberg 2012).
\bibitem{bertsch1980} G.F. Bertsch, The nuclear density of states in the space of nuclear shapes, Phys. Lett. {\bf B95}, 157 (1980).
\bibitem{barranco1990} F. Barranco, G.F. Bertsch, R.A. Broglia, and E. Vigezzi, Large-amplitude motion in superfluid Fermi droplets, Nucl. Phys. {\bf A512}, 253 (1990).
\bibitem{bertsch1994} G.F. Bertsch, Large amplitude collective motion, Nucl. Phys. {\bf A574}, 169c (1994).
\bibitem{PRL__2016} A. Bulgac, P. Magierski, K.J. Roche, and I. Stetcu, 
              Induced Fission of ${}^{240}$Pu within a Real-Time Microscopic Framework,
              Phys. Rev. Lett. {\bf 116}, 122504 (2016). 
\bibitem{BrinkBroglia}  D.M. Brink, R.A. Broglia, Nuclear Superfluidity: Pairing in Finite Systems, Cambridge University Press 2005. 
\bibitem{SM} See Supplemental Material at \{URL will be provided by
  the publisher\} for discussion of technical aspects including generation of
  initial configurations, derivation of energy of the junction, supplemental
  results, and list of movies, which includes Refs.~\cite{VortexPinning,MainBulgacPRL200,MainPRL__2002,MainPRC__2002,COCGMain,Bass1974,golabek,BlochMessiahA,ZuminoA,Mainlacroix,Mainebata}.
\bibitem{VortexPinning} G. Wlazłowski, K. Sekizawa, P. Magierski, A. Bulgac, M.M. Forbes, Vortex pinning and dynamics in the neutron star crust, Phys. Rev. Lett. {\bf 117}, 232701 (2016).
\bibitem{MainBulgacPRL200} Yongle Yu and Aurel Bulgac, %
   Energy Density Functional Approach to Superfluid Nuclei, %
   Phys. Rev. Lett. {\bf 90}, 222501 (2003).
\bibitem{MainPRL__2002} A. Bulgac and Y. Yu, %
  Renormalization of the Hartree-Fock-Bogoliubov Equations in the Case of a
  Zero Range Pairing Interaction, %
  Phys. Rev. Lett. {\bf 88}, 042504 (2002).
\bibitem{MainPRC__2002} A. Bulgac, %
  Local Density Approximation for Systems with Pairing Correlations, %
  Phys. Rev. C {\bf 65} 051305(R) (2002).
\bibitem{COCGMain} S.~Jin, A.~Bulgac, K.~Roche, and G.~Wlaz\l{}owski, 
Coordinate-Space Solver for Superfluid Many-Fermion Systems with 
Shifted Conjugate Orthogonal Conjugate Gradient Method, Phys. Rev. C {\bf 95}, 044302 (2017).  
\bibitem{Bass1974} R. Bass, Fusion of heavy nuclei in a classical model, Nucl. Phys. {\bf A231}, 45 (1974).
\bibitem{golabek} C. Golabek and C. Simenel, Collision dynamics of two 
$^{238}$U atomic nuclei, Phys. Rev. Lett. {\bf 103}, 042701 (2009).
\bibitem{BlochMessiahA} C. Bloch and A. Messiah, The canonical form of an antisymmetric tensor and its application to the theory of superconductivity, Nucl. Phys. {\bf 39}, 95 (1962).
\bibitem{ZuminoA} B. Zumino, Normal Forms of Complex Matrices, J. Math. Phys. {\bf 3}, 1055 (1962).
\bibitem{Mainlacroix} G. Scamps, D. Lacroix, G.F. Bertsch, and K. Washiyama, Pairing dynamics in particle transport, Phys. Rev. C \textbf{85}, 034328 (2012).
\bibitem{Mainebata} S. Ebata, T. Nakatsukasa, T. Inakura, K. Yoshida, Y. Hashimoto, and K. Yabana, Canonical-basis time-dependent Hartree-Fock-Bogoliubov theory and linear-response calculations, Phys. Rev. C \textbf{82}, 034306 (2010).
\bibitem{dietrich1} K. Dietrich, On a nuclear Josephson effect in heavy ion scattering, Phys. Lett. {\bf B32}, 428 (1970).
\bibitem{dietrich2} K. Dietrich, Semiclassical Theory of a Nuclear Josephson Effect in Reactions between Heavy Ions,
                    Ann. Phys. (N.Y.) {\bf 66}, 480, (1970).
\bibitem{dietrich3} K. Dietrich, K. Hara, and F. Weller, Multiple pair transfer in reactions between heavy nuclei, 
                    Phys. Lett. {\bf B35}, 201 (1971).
\bibitem{sorensen} J. H., Sorensen and A. Winther, Multipair transfer in collisions between heavy nuclei, Phys. Rev. C {\bf 47}, 1691 (1993).
\bibitem{hashimoto2016} Y. Hashimoto and G. Scamps, Gauge angle dependence in TDHFB calculations of 
$^{20}$O+$^{20}$O head-on collisions with the Gogny interaction, Phys. Rev. C {\bf 94}, 014610 (2016).
\bibitem{Oliveira} L. N. Oliveira, E. K. U. Gross, and W. Kohn, 
              Density-Functional Theory for Superconductors, 
              Phys. Rev. Lett. 60, 2430 (1988). 
\bibitem{Wacker} O. -J. Wacker, R. K\"{u}mmel, and E. K. U. Gross,
              Time-Dependent Density-Functional Theory for Superconductors,
              Phys. Rev. Lett. 73, 2915 (1994).
\bibitem{PRL__2009} A. Bulgac and S. Yoon,
              Large Amplitude Dynamics of the Pairing Correlations in a Unitary Fermi Gas,
              Phys. Rev. Lett. {\bf 102}, 085302 (2009).          
\bibitem{Science__2011}  A. Bulgac, Y.-L. Luo, P. Magierski, K.J. Roche, and Y. Yu, 
              Real-Time Dynamics of Quantized Vortices in a Unitary Fermi Superfluid, 
              Science {\bf 332}, 1288 (2011).  
\bibitem{LNP__2012} A. Bulgac, P. Magierski, and M.M. Forbes,
              The Unitary Fermi Gas: From Monte Carlo to Density
              Functionals, 
              in {\it BCS-BEC Crossover and the Unitary Fermi Gas}, 
              edited by W. Zwerger, Lecture Notes in Physics, Vol. {\bf 836}, pp 305-373 (Springer, Heidelberg, 2012).                  
\bibitem{PRL__2012} A. Bulgac, Y.-L. Luo, and K.J. Roche, 
              Quantum Shock Waves and Domain Walls in Real-Time Dynamics of a Superfluid Unitary Fermi Gas,
              Phys. Rev. Lett. {\bf 108}, 150401 (2012).            
\bibitem{ARNPS__2013} A. Bulgac,
             Time-Dependent Density Functional Theory and Real-Time Dynamics of Fermi Superfluids,
              Ann. Rev. Nucl. Part. Sci. {\bf 63}, 97 (2013).        
\bibitem{PRC__2011} I. Stetcu, A. Bulgac, P. Magierski, and K.J. Roche,
             Isovector Giant Dipole Resonance from 3D Time-Dependent Density Functional Theory 
             for Superfluid Nuclei,
             Phys. Rev. C {\bf 84}, 051309(R) (2011).   
\bibitem{PRL__2015} I. Stetcu, C.A.  Bertulani, A. Bulgac, P. Magierski, and K.J. Roche, 
              Relativistic Coulomb Excitation within Time-Dependent Superfluid Local Density Approximation,
              Phys. Rev. Lett. {\bf 114}, 012701 (2015). 
\bibitem{Mag2016} P. Magierski, Nuclear Reactions and Superfluid Time Dependent Density Functional Theory,
invited paper honoring Prof. Joachim Maruhn's retirement to be published as a chapter
in ``Progress of time-dependent nuclear reaction theory"
(ed. Yoritaka Iwata) in the ebook series: ``Frontiers in nuclear and particle physics"
(Bentham Science Publishers);  arXiv:1606.02225.
\bibitem{Fayans1} S.A. Fayans, Towards a universal nuclear density functional, JETP Letters {\bf 68}, 169 (1998).
\bibitem{Fayans2} S.A. Fayans, S.V. Tolokonnikov, E.L. Trykov, and D. Zawischa, Nuclear isotope shifts within the local energy-density functional approach, Nucl. Phys. {\bf A676}, 49 (2000).
\bibitem{arXiv:1702.00069} K. Sekizawa, P. Magierski, and G. Wlaz{\l}owski, Solitonic Excitations in Collisions of Superfluid Nuclei, arXiv:1702.00069 (2017).
\bibitem{nambu} Y. Nambu, Quasi-Particles and Gauge Invariance in the Theory of Superconductivity, Phys. Rev. {\bf 117}, 648 (1960).
\bibitem{goldstone} J. Goldstone, Field theories with $\ll$Superconductor$\gg$ solutions, Il Nuovo Cimento {\bf 19}, 154 (1961).
\bibitem{Nakatsukasa2016} T. Nakatsukasa, K. Matsuyanagi, M. Matsuo, and K. Yabana, Time-dependent density-functional description of nuclear dynamics, Rev. Mod. Phys. {\bf 88}, 045004 (2016).

\bibitem{Bulgac} We thank Aurel Bulgac for pointing it to us.

\bibitem{liang2012} J.F. Liang, C.J. Gross, Z. Kohley, D. Shapira, R.L. Varner, J.M. Allmond, A.L. Caraley, K. Lagergren, and P. E. Mueller, Fusion probability for neutron-rich radioactive-Sn-induced reactions, Phys. Rev. C {\bf 85} 031601, (2012).
\bibitem{Sahm1985} C.C. Sahm, H.G. Clerc, K.-H. Schmidt, W. Reisdorf, P. Armbruster, F.P. Hessberger, J.G. Keller, G. M\"{u}nzenberg, D. Vermeulen, Fusion probability of symmetric heavy, nuclear systems determined from evaporation-residue cross sections, Nucl. Phys. {\bf A441}, 316 (1985).
\bibitem{swiatecki1982} W.J. Swiatecki, The dynamics of the fusion of two nuclei, Nucl. Phys. {\bf A376}, 275 (1982).
\bibitem{swiatecki1982a} S. Bjornholm and W.J. Swiatecki, Dynamical aspects of nucleus-nucleus collisions, Nucl. Phys. {\bf A391}, 471 (1982).
\bibitem{donangelo1986} R. Donangelo, L.F. Canto, Studies of nucleus-nucleus collisons with a schematic liquid-drop model and one-body dissipation,
                                Nucl. Phys. {\bf A451}, 349 (1986).
\bibitem{Washiyama(2015)}
K.~Washiyama, Microscopic analysis of fusion hindrance in heavy nuclear systems,
Phys. Rev. C {\bf 91}, 064607 (2015).
\bibitem{Scamps(2013)}
G. Scamps and D. Lacroix, Effect of pairing on one- and two-nucleon transfer below the Coulomb barrier: A time-dependent microscopic description, Phys. Rev. C \textbf{87}, 014605 (2013).
\bibitem{Scamps(2015)}
G. Scamps and D. Lacroix, Effect of pairing on transfer and fusion reactions, EPJ Web of Conf. \textbf{86}, 00042 (2015).
\bibitem{Ebata(2015)}
S.~Ebata and T.~Nakatsukasa, Repulsive Aspects of Pairing Correlation in Nuclear Fusion Reaction, JPS Conf. Proc. \textbf{6}, 020056 (2015).
\bibitem{arXiv:1701.06683} A. Bulgac and S. Jin, Dynamics of Fragmented Condensates and Macroscopic Entanglement, Phys. Rev. Lett. (to be published), arXiv:1701.06683


\end{thebibliography}

\begin{thebibliography}{99}
\bibitem{VortexAAA}
G.~Wlaz{\l}owski, K.~Sekizawa, P.~Magierski, A.~Bulgac, and M.M.~Forbes,
Vortex Pinning and Dynamics in the Neutron Star Crust,
Phys. Rev. Lett. {\bf 117}, 232701 (2016).
\bibitem{SuppBulgacPRL200AAA} Yongle Yu and Aurel Bulgac, %
   Energy Density Functional Approach to Superfluid Nuclei, %
   Phys. Rev. Lett. {\bf 90}, 222501 (2003).
\bibitem{SuppPRL__2002AAA} A. Bulgac and Y. Yu, %
  Renormalization of the Hartree-Fock-Bogoliubov Equations in the Case of a
  Zero Range Pairing Interaction, %
  Phys. Rev. Lett. {\bf 88}, 042504 (2002).
\bibitem{SuppPRC__2002AAA} A. Bulgac, %
  Local Density Approximation for Systems with Pairing Correlations, %
  Phys. Rev. C {\bf 65} 051305(R) (2002).
\bibitem{COCGSuppAAA} S.~Jin, A.~Bulgac, K.~Roche, and G.~Wlaz\l{}owski, 
Coordinate-Space Solver for Superfluid Many-Fermion Systems with 
Shifted Conjugate Orthogonal Conjugate Gradient Method, Phys. Rev. C {\bf 95}, 044302 (2017).

\bibitem{SupBass1974AAA} R. Bass, Fusion of heavy nuclei in a classical model, Nucl. Phys. {\bf A231}, 45 (1974).

\bibitem{simenelAAA} C. Golabek and C. Simenel, Collision dynamics of two $^{238}$U atomic nuclei, 
Phys. Rev. Lett. {\bf 103}, 042701 (2009).

\bibitem{BlochMessiahAAA} C. Bloch and A. Messiah, The canonical form of an antisymmetric tensor and its application to the theory of superconductivity, Nucl. Phys. {\bf 39}, 95 (1962).
\bibitem{ZuminoAAA} B. Zumino, Normal Forms of Complex Matrices, J. Math. Phys. {\bf 3}, 1055 (1962).
\bibitem{lacroixSuppAAA} G. Scamps, D. Lacroix, G.F. Bertsch, and K. Washiyama, Pairing dynamics in particle transport, Phys. Rev. C \textbf{85}, 034328 (2012).
\bibitem{ebataSuppAAA} S. Ebata, T. Nakatsukasa, T. Inakura, K. Yoshida, Y. Hashimoto, and K. Yabana, Canonical-basis time-dependent Hartree-Fock-Bogoliubov theory and linear-response calculations, Phys. Rev. C \textbf{82}, 034306 (2010).

\end{thebibliography}
\end{document}